\documentclass[11pt,twocolumn]{aastex63}
\accepted{February 18, 2021}
\submitjournal{ApJ}

\shorttitle{3D Morphology of Open Clusters in the Solar Neighborhood}
\shortauthors{Pang et al.}

\begin{document}

\title{3D Morphology of Open Clusters in the Solar Neighborhood with Gaia EDR3: \\
        its Relation to Cluster Dynamics}

\author[0000-0003-3389-2263]{Xiaoying Pang}
    \affiliation{Department of Physics, Xi'an Jiaotong-Liverpool University, 111 Ren’ai Road, 
                Dushu Lake Science and Education Innovation District, Suzhou 215123, Jiangsu Province, P. R. China}
    \email{Xiaoying.Pang@xjtlu.edu.cn}
    \affiliation{Shanghai Key Laboratory for Astrophysics, Shanghai Normal University, 
                100 Guilin Road, Shanghai 200234, P. R. China}
                
\author{Yuqian Li}
    \affiliation{Department of Physics, Xi'an Jiaotong-Liverpool University, 111 Ren’ai Road, 
                Dushu Lake Science and Education Innovation District, Suzhou 215123, Jiangsu Province, P. R. China}

\author[0000-0001-6980-2309]{Zeqiu Yu}
    \affiliation{Department of Physics, Xi'an Jiaotong-Liverpool University, 111 Ren’ai Road, 
                Dushu Lake Science and Education Innovation District, Suzhou 215123, Jiangsu Province, P. R. China}

\author[0000-0003-4247-1401]{Shih-Yun Tang}
    \affiliation{Lowell Observatory, 1400 W. Mars Hill Road, Flagstaff, AZ 86001, USA}
    \affiliation{Department of Astronomy and Planetary Sciences, Northern Arizona University, Flagstaff, AZ 86011, USA}

\author[0000-0001-5532-4211]{Franti\v{s}ek Dinnbier}
    \affiliation{Astronomical Institute, Faculty of Mathematics and Physics, Charles University in Prague, V Hole\v{s}ovi\v{c}k\'{a}ch 2, 180 00 Praha 8, Czech Republic}

\author[0000-0002-7301-3377]{Pavel Kroupa}
    \affiliation{Astronomical Institute, Faculty of Mathematics and Physics, Charles University in Prague, V Hole\v{s}ovi\v{c}k\'{a}ch 2, 180 00 Praha 8, Czech Republic}
    \affiliation{Helmholtz-Institut f{\"u}r Strahlen- und Kernphysik (HISKP), Universität Bonn, Nussallee 14–16, 53115 Bonn, Germany}

\author[0000-0003-3784-5245]{Mario Pasquato}
    \affiliation{Center for Astro, Particle and Planetary Physics (CAP$^3$), New York University Abu Dhabi}
    \affiliation{INFN- Sezione di Padova, Via Marzolo 8, I–35131 Padova, Italy}

\author[0000-0002-1805-0570]{M.B.N. Kouwenhoven}
    \affiliation{Department of Physics, Xi'an Jiaotong-Liverpool University, 111 Ren’ai Road, 
                Dushu Lake Science and Education Innovation District, Suzhou 215123, Jiangsu Province, P. R. China}

%-------------------------------------------------------------------------------------------% 
%*******************************************************************************************%
\begin{abstract} 

We analyze the 3D morphology and kinematics of 13 open clusters (OCs) located within 500\,pc of the Sun, using {\it Gaia} EDR\,3 and kinematic data from literature. Members of OCs are identified using the unsupervised machine learning method \textsc{StarGO}, using 5D parameters ($X, Y, Z$, $\mu_\alpha \cos\delta, \mu_\delta$). The OC sample covers an age range of 25\,Myr--2.65\,Gyr. 
We correct the asymmetric distance distribution due to the parallax error using Bayesian inversion. The uncertainty in the corrected distance for a cluster at 500~pc is 3.0--6.3~pc, depending on the intrinsic spatial distribution of its members. We determine the 3D morphology of the OCs in our sample and fit the spatial distribution of stars within the tidal radius in each cluster with an ellipsoid model. The shapes of the OCs are well-described with oblate spheroids (NGC\,2547, NGC\,2516, NGC\,2451A, NGC\,2451B, NGC\,2232), prolate spheroids (IC\,2602, IC\,4665, NGC\,2422, Blanco\,1, Coma Berenices), or triaxial ellipsoids (IC\,2391, NGC\,6633, NGC\,6774). 
The semi-major axis of the fitted ellipsoid is parallel to the Galactic plane for most clusters. Elongated filament-like substructures are detected in three young clusters (NGC\,2232, NGC\,2547, NGC\,2451B), while tidal-tail-like substructures (tidal tails) are found in older clusters (NGC\,2516, NGC\,6633, NGC\,6774, Blanco\,1, Coma Berenices). Most clusters may be super-virial and expanding. $N$-body models of rapid gas expulsion with an SFE of $\approx 1/3$ are consistent with clusters more massive than $250\,\rm M_\odot$, while clusters less massive than 250\,$\rm M_\odot$ tend to agree with adiabatic gas expulsion models. Only six OCs (NGC\,2422, NGC\,6633, and NGC\,6774, NGC\,2232, Blanco\,1, Coma Berenices) show clear signs of mass segregation. 
\end{abstract}

\keywords{stars: evolution --- open clusters and associations: individual -- stars: kinematics and dynamics -- methods: statistical -- methods: numerical }

%-------------------------------------------------------------------------------------------%
%*******************************************************************************************%
\section{Introduction}\label{sec:intro}

%%%%%%%%%%%%%%%%%%%%%%%%%%%%%%%%%%%%%

Open star clusters (OCs) are stellar systems formed in giant molecular clouds (GMCs) located in the disk of the Milky Way \citep[e.g.,][]{lada2003}. Unlike their compact halo counterparts (globular clusters), the stellar members of OCs have a looser spatial distribution, hence the name ``open''. 
The formation and evolution of OCs is closely related to Galactic star formation. Studying the spatial distribution of stars in OCs therefore provides an opportunity to  uncover the mechanisms and conditions of star cluster formation in the Galaxy.

The earliest morphological study of OCs dates back a century \citep{jeans1916}. In the decades that followed, further systematic studies were carried out; notable studies include those of \citet{oort1979} and \citet{bergond2001}. They investigated the spatial distribution of a handful of nearby OCs, and found that the  flattening of the projected shape of OCs tends to be parallel to the Galactic plane. The pioneering work of \citet{chen2004} determined the 2D morphology of nearby 31 OCs using 2MASS infrared photometry; they took an important first step in the statistical investigation of OC morphology. However, they did not reach a firm quantitative conclusion due to the challenges arising from membership determination. 

Differences between the morphologies of young and old OCs were identified by \citep{jones2002}. Young OCs tend to have a higher degree of substructure.  \citet{Sanchez2009} found that clusters with fractal-like structures are generally younger than clusters with smooth radial density profiles. \citet{kounkel2019} identified several hundreds of filamentary structures younger than 100~Myr, most of which were associated with nearby OCs.  One string-like structure in \citet{kounkel2019}, for example, hosts two coeval open clusters, NGC\,2232 and LP\,2439 \citep{pang2020}. The extended substructures of young OCs are thought to have been inherited from the primordial shape of the parental GMCs \citep{ballone2020}, in which star formation takes place along the densest filamentary substructures \citep{jerabkova2019}.

Most regions in GMCs are not self-gravitating, and are supported by large-scale turbulence. Elongated shapes, such as triaxial and prolate shapes are therefore common among GMCs \citep{jones2002}. The triaxiality is consistent with the non-equilibrium state of GMCs. The dense cores in GMCs, where OCs are formed, are pulled together by self-gravity, with an observed elongated shape  \citep{curry2002}. After the first stars have formed, the gas surrounding the OCs is rapidly removed by stellar radiation \citep{krumkolz2009,dinnbier2020c}, stellar winds \citep{weaver1977}, and/or supernovae \citep{mckee1977}. Stars that escape from the cluster after gas expulsion reduce the gravitational potential of the cluster, and form a tidal ``tail~I'' \citep[following the definition and nomenclature of][]{dinnbier2020b}. Expansion has been observed in very young OCs with ages less than 5\,Myr \citep{kuhn2019}, as well as in young clusters that are tens of millions of years old \citep{brandner2008,bravi2018,getman2018, karnath2019, pang2020}.

Simultaneously, the stellar members of an OC interact with each other through two-body relaxation, which results in the observed “mass segregation” in star clusters \citep{hillenbrand1998,pang2013, tang2018}, in which low-mass stars are dispersed to the outskirts of the cluster and massive stars tend to migrate to the central region of the cluster. Consequently, a dense core will form, while low-mass stars continue to escape from the cluster, mainly at low speeds through Lagrange points \citep{kupper2008}, and form an S-shaped tidal ``tail~II'' \citep[following the nomenclature of][]{dinnbier2020b}. The reduction of cluster members further lowers the gravitational potential, which results in expansion of OCs and consequently a lower stellar number density. \citet{chen2004} proposed that the internal relaxation process causes the inner part of a cluster to evolve into a spherical spatial distribution.

As the Galactic disk is abundant in stars, spiral arms, and GMCs, OCs are subjected to external tidal perturbations, such as disk shocks, spiral arm passages, and encounters with molecular clouds \citep{spitzer1958,lamers2005,kruijssen2012}. Stars escape the cluster as a consequence of gas expulsion, close encounters or evaporation, and due to their exposure to the Coriolis force produced by the Galactic tidal field, and migrate to more tangential orbits. 
As a consequence, the star cluster stretches. 
Furthermore, when OCs cross the Galactic plane, the disk tidal field compresses them and flattens their shapes. The projected major axis of the elongated shapes of OCs are known to be aligned with the Galactic plane in most cases \citep{oort1979,bergond2001,chen2004}. As OCs evolve, their shapes continue to distort and members disperse, leading to the inevitable dissolution of the entire cluster. Expansion has been identified in old open clusters as a sign of an ongoing disruption process \citep{pang2018}. Giant tidal tails extending from OCs have been directly observed in recent years \citep{roser2019, meingast2019, tang2019, furnkranz2019, zhang2020}. These observed tidal tails are thought to be composed of both a ``tail I'', driven by gas expulsion, and by a ``tail II'', driven by evaporation \citep{dinnbier2020a, dinnbier2020b}. 
 
 The {\it Gaia} Early Data Release 3 \citep[EDR\,3;][]{gaia2020} has revolutionized the study of OC morphology by providing parallaxes with a 30\% higher precision and proper motions with double accuracy, as compared to those in the {\it Gaia} Data Release 2 \citep[DR\,2;][]{gaia2018a}. It is  desirable to represent the stellar distribution of OCs in three dimensions in order to reveal the formation process and early evolution of OCs. Besides, it is necessary that this kind of analysis is carried out by parameterizing the cluster shape in an objective, quantitative, as well as systematic manner. 

In this study, we conduct a statistical analysis of the morphology of 13~OCs located within 500\,pc from the Sun (see Table~\ref{tab:general}) in the solar neighborhood based on {\it Gaia} EDR\,3 data. The distances to the target clusters range between $\approx$86\,pc (Coma Berenices) and $\approx$476\,pc (NGC\,2422). The target OCs span a representative range in ages, from $\approx$25\,Myr (NGC\,2232) to $\approx$2.65\,Gyr (NGC\,6774). Among the OCs in this study, three clusters have membership determination that was carried out in previous works: Coma Berenices \citep{tang2019}, Blanco~1 \citep{zhang2020}, and NGC\,2232 \citep{pang2020}. We are motivated to quantify the shapes of the clusters in the sample, and establish their relation to the dynamical state of each of the OCs, which is quantified with kinematic data from the literature. The present study is a pioneering work setting up the tools to quantify 3D morphology of OCs using {\it Gaia} EDR\,3 data. At the same time, it is also a reminiscent analogy to the studies that quantified the morphology of elliptical galaxies \citep{benacchio1980,padilla2008}.

The paper is organized as follows. In Section~\ref{sec:gaia_member} we discuss the quality and limitations of the {\it Gaia} EDR\,3 data, and describe our input data-set for member star identification. We then present the algorithm, \textsc{StarGO}, which is used to determine cluster members. The properties of the identified member candidates of the 13 target OCs are discussed in Section~\ref{sec:result}. The 3D morphology of target OCs and the parameterization of the cluster shapes are presented in Section~\ref{sec:3D}, in which we reconstruct the distances with a Bayesian method  (Section~\ref{sec:dis_correct}). The dynamical state of the OCs are quantified using kinematic data in Section~\ref{sec:dynamics}. In Section~\ref{sec:nbody} we compare our observational findings with $N$-body simulations. Finally, we provide a brief summary in Section~\ref{sec:summary}.

%-------------------------------------------------------------------------------------------%
%-------------------------------------------------------------------------------------------%
\section{Data Analysis and Member identification}\label{sec:gaia_member}

%-------------------------------------------------------------------------------------------%
\subsection{Gaia EDR\,3 Data Processing and Analysis}\label{sec:target_selection}

The {\it Gaia} EDR\,3 \citep{gaia2020} has provided parallaxes ($\varpi$) and proper motions (PMs; $\mu_\alpha \cos\delta, \mu_\delta$) with unprecedented precision and sensitivity for more than 1.8 billion sources  with magnitude brighter than 21~mag in the $G$~band ($330-1050$~nm). The uncertainty in the $G$~band photometry is in the range of 0.2--6~mmag for stars brighter than 20~mag. The median error of $\varpi$ ranges from $\sim$0.02--0.03~mas for bright sources ($G$\,$<$\,15~mag) to $\sim$1.3~mas for faint stars ($G$\,$\approx$\,21~mag). The corresponding uncertainties in the PMs for these sources are 0.02--0.03\,mas\,yr$^{-1}$ and 1.4\,mas\,yr$^{-1}$, respectively \citep{gaia2020}. Beside PMs, 
about 7.2~million stars have radial velocity (RV) measurements in the {\it Gaia} DR\,2, which are transferred to EDR\,3 \citep{torra2020,seabroke2020}. These RV measurements have a typical uncertainty of 2~$\rm km\,s^{-1}$ \citep{lindegren2018}. Unreliable or erroneous RVs in the data release have been discarded \citep[see][]{boubert2019}.  

The following analysis is carried out for these 13 OCs in the the sample.
The spatial and kinematic structures of the ten target clusters are investigated using {\it Gaia} EDR\,3 data within 100\,pc from the cluster center taken from the member catalogs of \citet{liu2019} and \citet{gaia2018b} in Cartesian Galactocentric coordinates (see definition in Appendix~\ref{sec:apx_coordinate}).
In order to remove possible artifacts in the {\it Gaia} EDR\,3 from our sample, we apply a general astrometric quality cut, as described in \citet[in their Appendix~C]{lindegren2018}, which selects stars with parallaxes and photometric measurements within 10 percent uncertainty. Hereafter, we refer to this set as ``Sample~I''. 
Generally, the number of stars in Sample~I ranges from  122\,154 to 456\,527 for the clusters in our study. The $G$-band magnitude of the sources in Sample~I ranges between $\sim$3.0\,mag and $\sim$20.7\,mag. For most clusters in the sample, measurements become significantly incomplete for $G \ga 19$~mag. 

\begin{figure*}[tb!]
\centering
\includegraphics[angle=0, width=1.\textwidth]{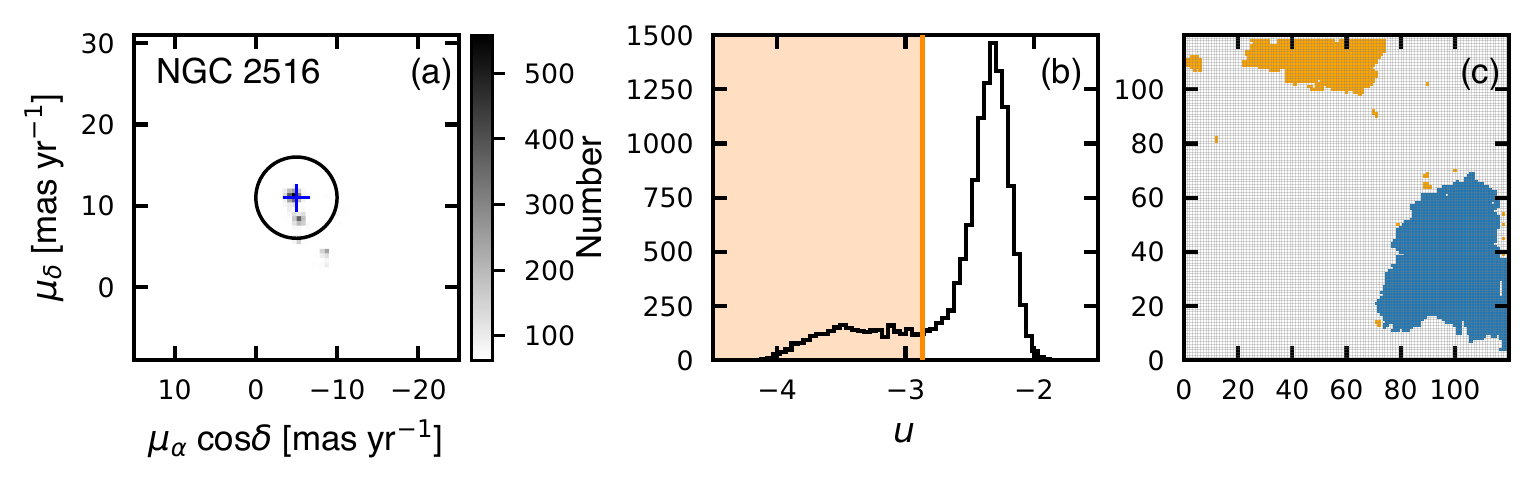}
\caption{ 	 
    (a)~ 2D density map of the proper motion vectors for the regions around NGC\,2516 in sample~I. The blue cross indicates the mean over-density generated by NGC\,2516 obtained from \citet{liu2019}. Each bin is smoothed by neighboring eight bins. Only bins with a 
        number count $>3\sigma$ are shown, where $\sigma$ is the standard deviation of all bins.
        The color indicates the number count in each bin.
	(b)~ Histogram of the distribution of $u$. 
	     The orange line denotes the selections of $u$ that produces a 5\%
	    contamination rate among the identified candidates, for the orange patch
	    in the 2D neural network (panel (c)).
	(c)~ 2D neural network} resulting from SOM, the neurons with a $u$-selection of 5\% contamination rate  (orange line in panel (b)) are shaded as orange. Among these, the neurons corresponding to the member candidates of the target cluster NGC\,2516 are highlighted in blue.
\label{fig:som}
\end{figure*}

We construct a 2D density map of PMs to select stars around the over-density location of the 13 target clusters. Figure~\ref{fig:som}~(a) shows a 2D density map of PMs for the cluster NGC\,2516 as an example. This map only shows bins with over-densities $>3\sigma$ in Sample~I. Several over-densities stand out. There is an over-density near the average PMs of the cluster (indicated with a blue cross) provided by \citet{liu2019}. An over-density of nearby clusters can also be seen in Figure~\ref{fig:som} (PM plots for other twelve OCs are provided in Appendix~\ref{sec:apx_som}; see Figures~\ref{fig:som_a1}, \ref{fig:som_a2} and \ref{fig:som_a3}). 
In this work we focus on the target clusters and we do not investigate their neighbors. We apply a circular cut (the black circle in Figure~\ref{fig:som}~(a)) to only include the target cluster for further analysis. Note that the radius of the circle is chosen to include as many potential members as possible, while simultaneously excluding most unrelated nearby structures. The radius is therefore different for each cluster in the sample. 
Application of this circular cut reduces the number of candidate members for each cluster. Hereafter, we refer to this set of stars as ``Sample~II''. The number of stars in Sample~II drops to below 10\,000 for most clusters. The stars in this sample have magnitudes ranging between $G \sim 3.8$~mag and $G \sim 20.6$~mag. All samples are complete for $G\la 18-18.5$~mag. 

%fig1

In this study we use the 5D parameters of stars in Sample~II 
(R.A., Decl., $\varpi$, $\mu_\alpha \cos\delta$, and $\mu_\delta$) from {\it Gaia} EDR\,3. Since only a small fraction of the stars in each cluster have RV measurements, we adopt the higher-accuracy radial velocities from \citet{jackson2020} and \citet{bailey2018} as supplementary data. The RVs of stars in the ten target clusters are obtained from \citet{jackson2020}; these are part of the Gaia-ESO Survey \citep[GES,][]{gilmore2012} with an uncertainty of 0.4~km\,s$^{-1}$, obtained using FLAMES (the Fiber Large Array Multi Element Spectrograph) combined with the GIRAFFE
and UVES (Ultraviolet and Visual Echelle Spectrograph) spectrographs mounted on the 8-m UT2-Kueyen telescope of the ESO Very Large Telescope facility. 
\citet{bailey2018} obtained RVs of stars in 
NGC\,2422 with M2FS (the Michigan/Magellan Fiber System), a multi-object fibre-fed spectrograph on the Magellan/Clay 6.5-m telescope, with a median uncertainty of 0.08~km\,s$^{-1}$. We use the RVs from Gaia DR\,2 for the clusters Coma Berenice and NGC\,6774, neither of which is included in the above-mentioned spectroscopic surveys. 

The distance to each individual star is computed as $1/\varpi$, from which we compute for each source the Galactocentric Cartesian coordinates ($X, Y, Z$). The transformation is performed by using the Python \texttt{Astropy} 
package \citep{astropy2013, astropy2018}. There is an asymmetric error in the distance that arises from the direct
inversion of $\varpi$ \citep{zhang2020}. We adopt 
a Bayesian method to correct individual distances of stars, as outlined in Section~\ref{sec:3D}.

%-------------------------------------------------------------------------------------------%
\subsection{Membership determination}\label{sec:stargo}

The unsupervised machine learning method,
\textsc{StarGO} \citep{yuan2018}\footnote{\url{https://github.com/salamander14/StarGO}} has proven to be successful in membership determination of OCs, e.g., for the Coma Berenices cluster \citep{tang2019}, 
Blanco\,1 \citep{zhang2020}, NGC\,2232 and LP\,2439 \citep{pang2020}. 
The algorithm is based on the Self-Organizing-Map (SOM)  
method that maps high-dimensional data onto a two-dimension neural network, while preserving the topological structures of the data. 

We apply \textsc{StarGO} to map a 5D data set ($X, Y, Z$, $\mu_\alpha \cos\delta, \mu_\delta$) of ten target clusters (Sample~II) onto a 2D neural network in order to determine member candidates. 
Stars are fed to the neural network sequentially. We therefore scale
the number of neurons to the number of stars in Sample~II. We adopt a network with 100$\times$100--150$\times$150 neurons (depending on the number of stars in Sample~II of each cluster) represented by the 100$\times$100 (150$\times$150) grid elements to study Sample~II (an illustration for NGC\,2516 is provided in Figure~\ref{fig:som}~(c)). Each neuron is assigned a random 5D weight vector with the same dimensions as the observed 5D parameters ($X, Y, Z$, $\mu_\alpha \cos\delta, \mu_\delta$) that are provided to the algorithm. During each iteration, the weight vector of each neuron is updated so that it is closer to the input vector of an observed star. The learning process is iterated 400 times (600 times for 150$\times$150 grids) until the weight vectors converge.
When stars associated with neurons are spatially and kinematically coherent (e.g., when they are cluster members), the 5D weight vectors of the adjacent neurons are similar. Therefore, the value of the difference of weight vectors between these adjacent neurons, $u$, is small.  
Neurons with similar small values of $u$ group together in the 2D neural network as patches (see Figure~\ref{fig:som}~(c)). Different groups of stars form different patches. The value of $u$ is smaller for 
neurons located inside the patch, and larger for neurons outside the patch. The $u$ values of neurons inside patches generate an extended tail towards small values in the $u$-histogram (see panel (b) in Figure~\ref{fig:som}).

The selection of $u$ is made by applying a cut to the tail of the $u$-distribution. This cut is made to ensure a similar contamination rate of $\sim$5\% among members, which has been applied to NGC\,2232 in \citet{pang2020}. We adopt this 5\% field star contamination $u$-cut as a member selection criteria for the ten target clusters, which corresponds to the blue patch in Figure~\ref{fig:som}~(c). We evaluate the contamination rate from the smooth Galactic disk population using the {\it Gaia} DR\,2 mock catalog \citep{rybizki2018}. An identical PM cut as described in Section~\ref{sec:target_selection} is also applied to the mock catalog in the same volume of the sky. Each of these mock stars is attached to the trained 2D neural network. We then consider the mock stars associated with selected patches as contamination. The numbers of identified members of each target cluster are listed in Table~\ref{tab:general}. We provide a  detailed member list of all 13 target clusters in Table~\ref{tab:memberlist}, with parameters obtained in this study. The  members lists of these 13 target clusters therefore form homogeneous data sets.

%-------------------------------------------------------------------------------------------%
\section{General properties of target open clusters}\label{sec:result}

To evaluate the validity our membership identification, we cross-match the members in target clusters with two independently published catalogs that both identify star clusters using all-sky Gaia DR\,2 data: \citet{liu2019} and  \citet{cantat2020}. \citet{liu2019} used a friend-of-friend (FoF) cluster finder to identify star clusters in {\it Gaia} DR\,2 in the five-dimensional parameter space ($l, b, \varpi, \mu_\alpha\cos\delta$, and $\mu_\delta$). Members in the catalog of \citet{cantat2020} are compiled from \citet{cantat2020a,castro2018,castro2019,castro2020} and are identified using the unsupervised membership assignment code \textsc{UPMASK} \citep{cantat2018}.  

All of the target clusters presented in this work are generally in good agreement with both catalogs, and have a comparable number of identified members (see the last two columns in Table~\ref{tab:general}). Coma Berenices, Blanco~1, and NGC\,6774 are absent in \citet{liu2019}'s catalog.

\begin{figure*}[tb!]
\centering
\includegraphics[angle=0, width=1.\textwidth]{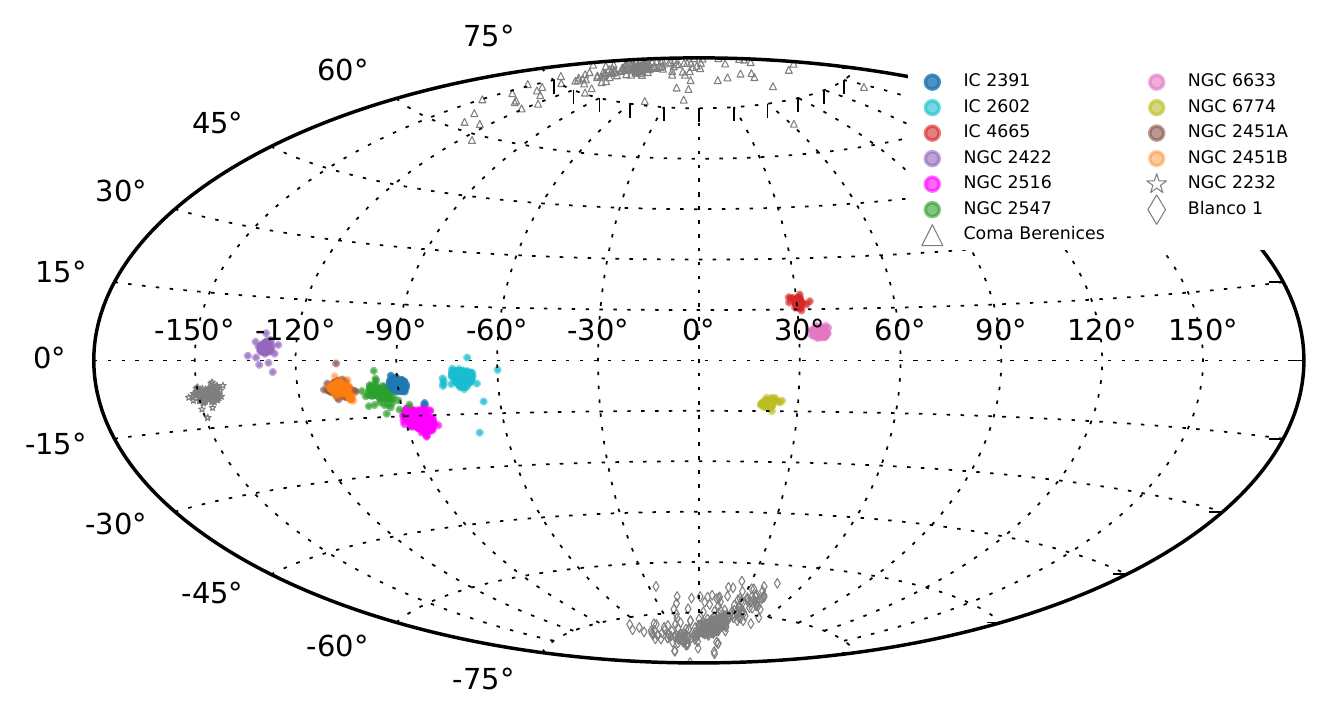}
\caption{
        2D projection of identified member stars of each target cluster in Galactic coordinates ($l, b$). Each of the 13 cluster for which the members are obtained via {\it Gaia} EDR\,3 in this study are denoted with different colours and symbols. Among these, three clusters are colored indicated in grey. The members of these clusters were also identified using {\it Gaia} DR\,2 in earlier studies. Member stars of NGC\,2232 are indicated with grey stars, those of Coma Berenices with grey triangles, and those of Blanco\,1 with grey diamonds. 
	    }
\label{fig:lb}
\end{figure*}

We display the positions of all identified members of the 13 target clusters in the Galactic coordinates in Figure~\ref{fig:lb}. Coma Berenices (grey triangles) and Blanco~1 (grey diamonds) occupy the regions of the Northern and Southern Galactic poles, respectively. The other OCs are within $\sim$15 degrees from the Galactic plane. Although NGC\,2451A and NGC\,2451B appear to overlap in the 2D projection, they are separated by a distance of $\approx 200$\,pc along the line-of-sight (see Table~\ref{tab:general}). Extended tidal tails are clearly visible in Coma Berenices and Blanco~1. An elongated shape is observed in the other clusters, notably in NGC\,2547, NGC\,2516, NGC\,2232 and NGC\,2451B. Note that a 2D elongated projected morphology that we see in projection, must have an even more prominent elongation in its 3D morphology. We will carry out an detailed investigation of the 3D morphology of the clusters in our sample in Section~\ref{sec:3D}.

%fig3

\begin{figure*}[tb!]
\centering
\includegraphics[angle=0, width=0.85\textwidth]{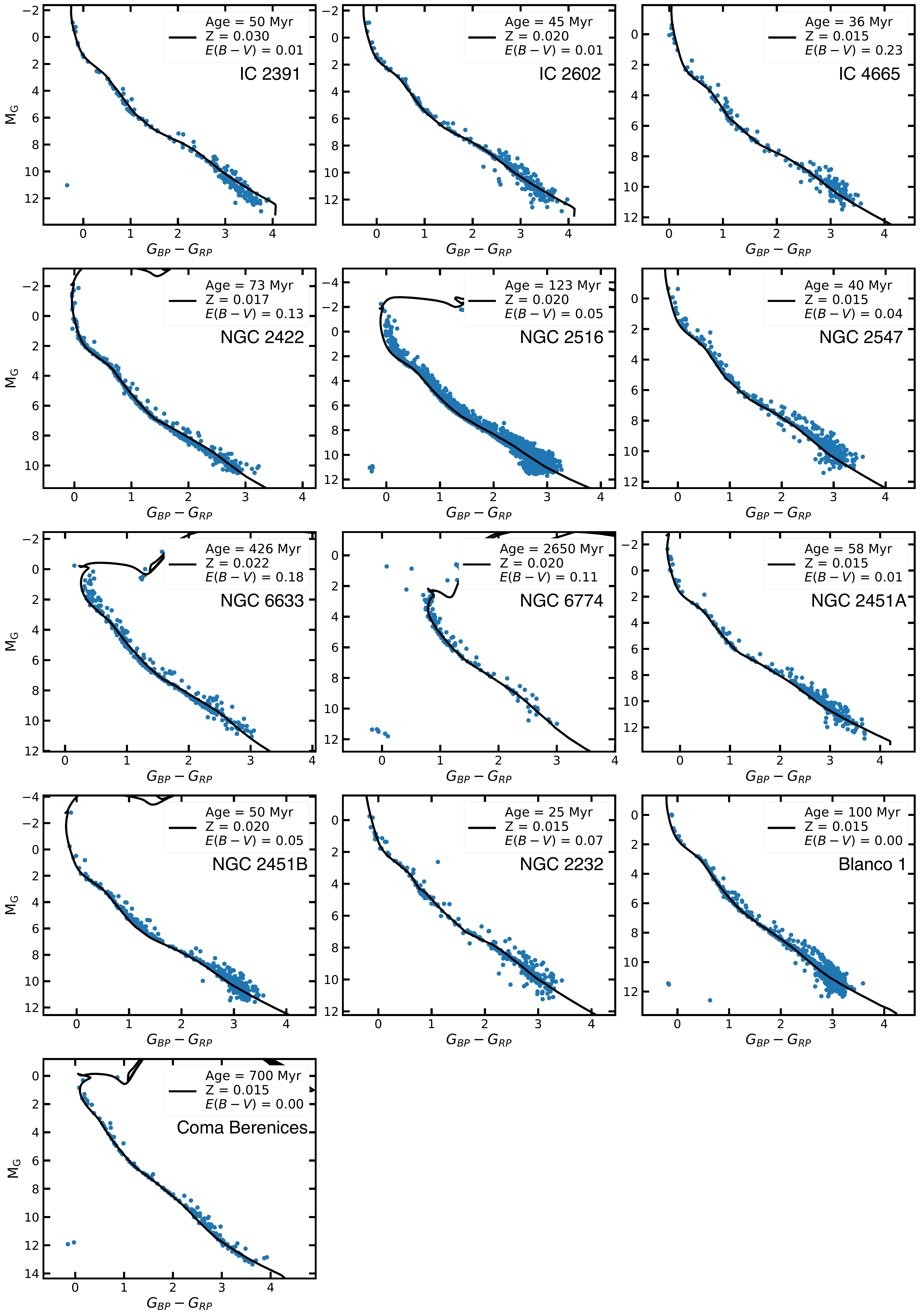}
\caption{
        The color-magnitude diagrams obtained from the {\it Gaia} EDR\,3 absolute magnitude M$_{G}$ 
        (adopting the distance after the correction described in Section~\ref{sec:dis_correct}) for member stars (blue dots) of 13 target OCs identified by \textsc{StarGO}. 
        The PARSEC isochrones of the adopted/fitted age are indicated with the black solid curves, with $E(B-V)$ and metallicities provided by literature or estimated in work (Table~\ref{tab:general}).  
	    }
\label{fig:iso}
\end{figure*}

We show the members of each cluster in the color-magnitude diagram (CMD; Figure~\ref{fig:iso}). The member stars of each cluster track a clear locus of a main sequence, which is consistent with the PARSEC isochrone (black solid curves in Figure~\ref{fig:iso}) for which the sensitivity curves are provided by \citet{mazapellaniz2018}. The distribution of the stars in the CMD shows that the field stars' main sequence (which is bluer than that of the clusters) is largely filtered out, therefore further confirming the reliability of our identified members of each cluster. We adopt ages for the target clusters from previous studies (except for NGC\,2451B) when they are in good agreement with the locations of the members in the CMD. We fit the values of $E(B-V)$ and of the metallicity, which are not available from literature.
We list the cluster ages and related  parameters in Table~\ref{tab:general}. The ages of the clusters span a wide range, from 25\,Myr for the youngest cluster (NGC\,2232) to 2.65\,Gyr for the oldest cluster (NGC\,6774). Such a wide age range in the sample of target clusters allows us to probe the influence of the secular dynamical evolution of the star clusters and the interaction with their environments on their morphology. The majority of the clusters in the sample are relatively young, with ages younger than 100\,Myr. Four clusters are of intermediate-age, with ages between 100\,Myr and 800\,Myr.  

An extended main sequence turn-off (eMSTO) of $\sim$0.3\,mag in the color $G_{BP}-G_{RP}$ is observed in two intermediate-age clusters, NGC\,2516 (123\,Myr) and NGC\,6633 (426\,Myr). The eMSTO region has been observed in many other star clusters \citep{li2014,li2017, milone2018,li2019}, which is a result of stars with a wide distribution of rotation rates \citep{bastian2009, dantona2017}. At the same time, the binary sequence locus of equal-mass systems is clearly seen for most clusters. In the oldest cluster NGC\,6774, we observe blue straggler candidates. 

Sixteen white dwarf members are found in five of the target clusters (IC\,2391, Blanco\,1, NGC\,2516, Coma Berenices, and NGC\,6774). 
The majority of these have been cataloged in \citet{fusillo2019}. The white dwarfs in NGC\,2516 and NGC\,6774 gather at very similar locations in the CMD. A detailed study has been carried out for three of the white dwarfs in NGC\,2516 by \citet[][IDs: NGC\,2516-1,2,5]{koester1996}. These white dwarfs were estimated to have ages of 120--160\,Myr (based on the cooling age, the main sequence lifetime, and the lifetime of red giants), which is consistent with the age of the cluster determined in our study. 

%fig2

% Table 1
\begin{deluxetable*}{cc rccR c RR LL rrr}%{r CLC c CLC}
\tablecaption{General parameters of target clusters \label{tab:general}
             }
% \tablewidth{\dimen1000}
\tabletypesize{\scriptsize}
\tablehead{
	 \colhead{Cluster}      & \colhead{Age} & 
	 \colhead{$Dist_{cor}$} & \colhead{$erDist_{cor}$}    & \colhead{$r_{\rm h}$}   & \colhead{$r_t$}   & 
	 \colhead{}             &
	 \colhead{$M_{cl}$}     & \colhead{$M_{dyn}$}    & \colhead{$Z$} & \colhead{$E(B-V)$}            &
	 \colhead{memb.}        & \colhead{CG20}        & \colhead{LP19}                \\
%----------------
	 \colhead{}             & \colhead{(Myr)}       &
	 \multicolumn{4}{c}{(pc)} &
	 \colhead{}             &
	 \multicolumn{2}{c}{(M$_\sun$)}   & 
	 \colhead{(dex)}      &  \colhead{(mag)}      & 
	 \multicolumn{3}{c}{(number)} \\
	 \cline{3-6} \cline{8-9} \cline{12-14} 
	 \colhead{(1)} & \colhead{(2)} & \colhead{(3)} & \colhead{(4)} & \colhead{(5)} & \colhead{(6)} && \colhead{(7)} &
     \colhead{(8)} & \colhead{(9)} & \colhead{(10)} & \colhead{(11)} & \colhead{(12)} & \colhead{(13)}
	 }
\startdata
IC\,2391    & 50$^{a,C}$    & 151.5    & 0.8   & 2.5   &  7.6  &&  140.2   & 315.8     & 0.030$^W$     & 0.01$^{b,C}$  & 219   & 190 (86\%)       & 135 (94\%)\\
IC\,2602    & 45$^{a,C}$    & 151.9    & 0.7   & 3.7   &  8.4  &&  188.1   & 464.8     & 0.020$^W$     & 0.01$^W$      & 318   & 267 (86\%)       & 135 (94\%)\\
IC\,4665    & 36$^{c,C}$    & 347.3    & 2.8   & 6.0   &  7.9  &&  158.5   & 530.5     & 0.015$^W$     & 0.23$^W$      & 197   & 142 (85\%)       & 74 (91\%)\\
NGC\,2422   & 73$^{d,C}$    & 476.5    & 3.2   & 4.9   & 11.4  &&  480.2   & 1112.4    & 0.017$^{d,C}$ & 0.13$^W$      & 466   & 312 (75\%)       & 335 (68\%)\\
NGC\,2516   & 123$^{e,C}$   & 410.5    & 3.1   & 7.9   & 18.3  && 1973.3   & 3368.2    & 0.020$^{e,C}$ & 0.05$^{e,C}$  & 2690  & 640 (98\%)       & 1365 (96\%)\\
NGC\,2547   & 40$^{e,C}$    & 387.4    & 2.7   & 5.6   &  9.8  &&  303.9   & 1032.2    & 0.015$^{e,C}$ & 0.04$^{e,C}$  & 452   & 192 (89\%)       & 214 (88\%)\\
NGC\,6633   & 426$^{f,C}$   & 394.3    & 2.4   & 5.3   & 10.2  &&  337.3   & 1698.2    & 0.022$^W$     & 0.18$^W$      & 300   & 133 (89\%)       & 164 (88\%)\\
NGC\,6774   & 2650$^{g,C}$  & 306.5    & 2.0   & 4.7   &  7.8  &&  152.6   & 1039.0    & 0.020$^{g,C}$ & 0.11$^{g,C}$  & 154   & 136 (80\%)       & \nodata\\ 
NGC\,2451A  & 58$^{h,C}$    & 192.6    & 1.3   & 4.9   &  8.3  &&  182.2   & 738.5    & 0.015$^W$     & 0.01$^{h,C}$  & 311   & 266 (80\%)       & 204 (93\%)\\
NGC\,2451B  & 50$^W$        & 362.9    & 2.6   & 6.0   &  9.1  &&  242.1   & 566.7    & 0.020$^W$     & 0.05$^{h,C}$  & 359   & 207 (73\%)       & 109 (85\%)\\
NGC\,2232   & 25$^{i,C}$    & 319.1    & 2.3   & 6.8   &  8.6  &&  205.8   & 263.7    & 0.015$^{i,C}$ & 0.07$^{i,C}$  & 281   & 169 (90\%)       & 93 (89\%)\\
Blanco\,1   & 100$^{j,C}$   & 236.7    & 2.1   & 6.7   & 10.2  &&  342.9   & 605.0    & 0.015$^{j,C}$ & 0.0$^{j,C}$   & 703   & 369 (97\%)       & \nodata\\
Coma Berenices    & 700$^{k,C}$   & 86.4     & 0.3   & 4.7   &  6.8  &&  101.6   & 574.0     & 0.015$^{k,C}$ & 0.0$^{k,C}$   & 158   & 129 (84\%)       & \nodata\\
\enddata
\tablecomments{ $Dist_{cor}$ is the mean corrected distance of members in each cluster. $erDist_{cor}$ is the error in corrected distance following the Bayesian model described in Section~\ref{sec:dis_correct}. $r_{\rm h}$ and $r_t$ are half-mass and tidal radii of each cluster. The metallicity, $Z$, and reddening, $E(B-V)$, of several clusters are taken from literature (indicated with a capital $C$), some are fitted in this work indicated with a capital $W$. When the referenced age fits the members, we adopt the age from previous works (indicated with a capital $C$).
 The quantity ${M}_{cl}$ is the mass of each star cluster. The last two columns show the number of matched members in \citet{cantat2018} (CG18) and \citet{liu2019} (LP19),and the corresponding percentages. The age, $Z$ and $E(B-V)$ for some clusters are adopted from 
a:\citet{marsden2009}; b:\citet{postnikova2020}; 
c: \citet{miret2019}, 
d: \citet{bailey2018}, e:\citet{gaia2018a}, f:\citet{williams2007}, 
g: \citet{olivares2019}, 
h: \citet{balog2009}, i:\citet{pang2020}, 
j: \citet{zhang2020}, 
k: \citet{tang2019}. 
    }
\end{deluxetable*}

%-------------------------------------------------------------------------------------------%

%-------------------------------------------------------------------------------------------%
\section{3D morphology of open clusters}\label{sec:3D}
\subsection{Distances through Bayesian parallax inversion}\label{sec:dis_correct}

It is known that the morphology of star clusters appears to be stretched along the line-of-sight, when distances are obtained by simple parallax inversion \citep[see, e.g.,][]{carrera2019}.
Such artificial elongation is a consequence of computing the distance to each star by directly inverting the Gaia EDR\,3 or DR\,2 parallax, $1/\varpi$. Even when the errors in the parallax measurements $\Delta\varpi$ have a symmetric distribution, taking the reciprocal introduces a skewed distribution of errors on the distances, which results in a systematic bias in the distance to each cluster.  We perform Monte Carlo simulations to estimate the contribution of the parallax error $\Delta \varpi$ on the uncertainty in the cluster distance \citep[see also][]{zhang2020}.
The mean value of $\Delta \varpi$ of each cluster is adopted to estimate the parallax induced uncertainty in the distance, which is typically 0.4--12.6\,pc for the clusters in our sample.
%and are listed in Table~\ref{tab:general}. 

To mitigate this issue, we follow the method introduced by \citet{bailer2015} and treat the inversion problem within a Bayesian framework. Our approach closely follows the distance correction procedure described in \citet{carrera2019}. In this approach a prior distribution is assumed for each star. The Bayesian theorem is adopted to estimate the prior through the likelihood function computed from the observed parallax and its nominal error. The prior is composed of two components, one representing the star cluster density and the other representing the field. The former follows a normal distribution, the latter an exponentially decreasing density as in \citet{bailer2015}. The standard deviation of the cluster component coincides with the standard deviation of cluster-centric distance of the member stars (i.e., the distances of the stars from the center of each star cluster). We combine these two components with weights proportional to the membership probability. We apply a value of 95\% for the cluster term, and 5\% for the field star term (see Section~\ref{sec:stargo}). The mean of the posterior is considered as the corrected distance for each star. Further details about this parallax inversion approach can be found in \cite{carrera2019} and in \cite{pang2020}.

\begin{figure*}[tb!]
\centering
\includegraphics[angle=0, width=1.\textwidth]{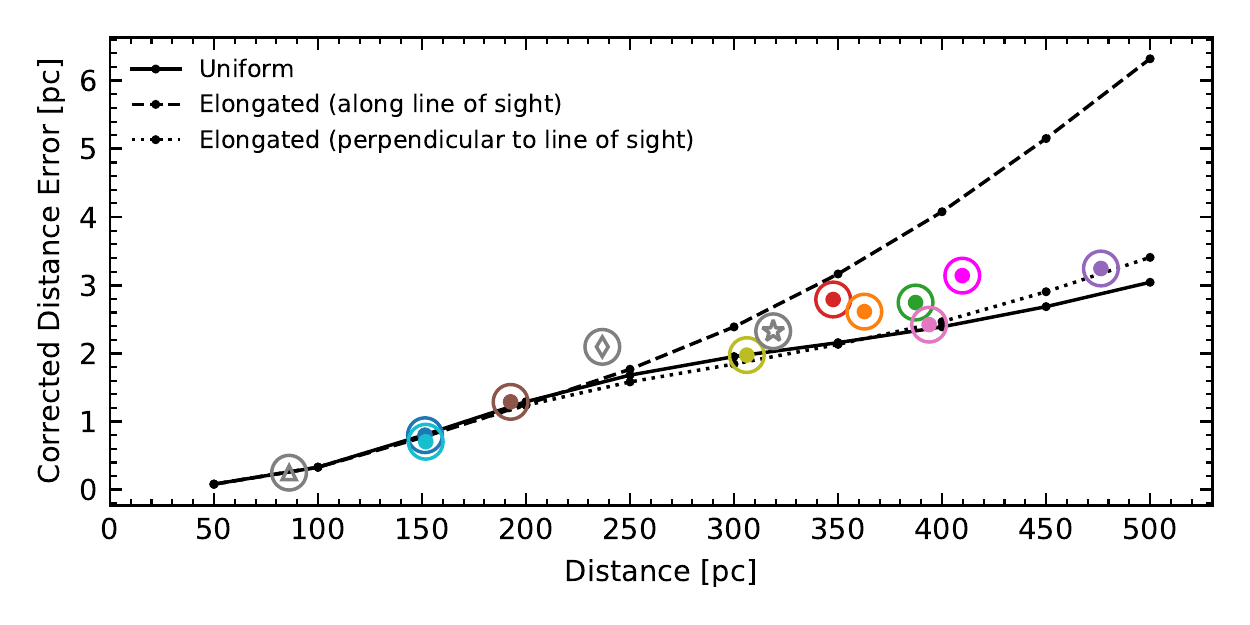}
\caption{
        Dependence of the uncertainty in the corrected distance on cluster distances based on simulations described in Appendix~\ref{sec:dis_correct_er}. The black solid curve represents a star cluster in which the members have a uniform spatial distribution. The dotted and dashed curves are clusters with elongated shape perpendicular to and parallel to the line-of-sight, respectively. The colored solar symbols and grey symbols indicate errors in the corrected distances when adopting a star cluster with a uniform stellar density located at the distance of each of the clusters in our study. The color coding of each cluster is identical to that in Figure~\ref{fig:lb}.
	    }
\label{fig:simulation_dis_er}
\end{figure*}

Furthermore, we carry out Monte-Carlo simulations to estimate the uncertainty in the corrected distances that result from our procedure. Three types of clusters are simulated to test our Bayesian procedure: (i) spherical star clusters with a uniform spatial distribution of stars; (ii) elongated star clusters with an elongation perpendicular to the line-of-sight; and (iii) elongated star clusters with an elongation along the line-of-sight. For the uniform model, the uncertainty in our corrected distance increases monotonically with distance (solid black curve in Figure~\ref{fig:simulation_dis_er}). At a distance of 500\,pc, the error in the mean corrected distance of all stars becomes as large as 3.0\,pc. When the elongation of the cluster is perpendicular to the line-of-sight, errors are very similar to those of a uniform cluster, and reach a slightly larger error of 3.4\,pc at a distance of 500\,pc (dotted black curve in Figure~\ref{fig:simulation_dis_er}). The situation is different for the model in which the elongation is along the line-of-sight; in this case the uncertainty is as large as 6.3\,pc at a distance of 500\,pc (dashed black curve in Figure~\ref{fig:simulation_dis_er}). Details of the procedure of the simulations are described in  Appendix~\ref{sec:dis_correct_er}. 

These findings show that the quality of the {\it Gaia} parallaxes play an important role in determining the recovered intrinsic cluster morphology from measurements. Artificial morphological elongation due to parallax errors becomes most severe when the intrinsic elongation happens to align with the line-of-sight. Intrinsically elongated clusters will suffer from larger uncertainty in the corrected distance, especially when their elongation is aligned with the line-of-sight.

%fig4

%-------------------------------------------------------------------------------------------%

%-------------------------------------------------------------------------------------------%
\subsection{Presentation of 3D morphology}\label{sec:3D_inte}

\begin{figure*}[tb!]
\centering
\includegraphics[angle=0, width=\textwidth]{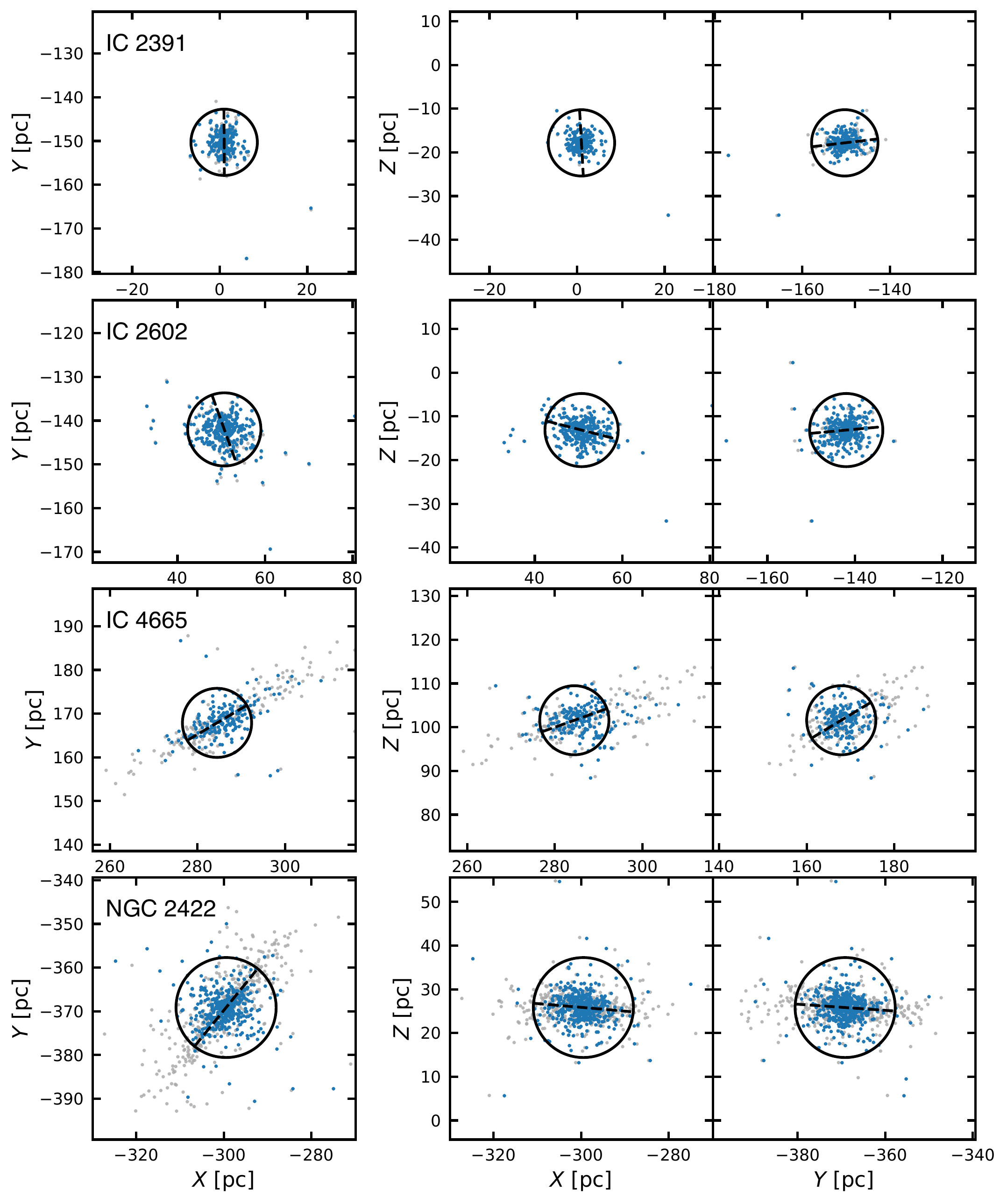}
\caption{3D spatial position of member stars in four target clusters: IC\,2391, IC\,2602, IC\,4665, NGC\,2422, 
        in heliocentric Cartesian coordinates ($X,Y,Z$; see definition in  Appendix~\ref{sec:apx_coordinate})            
        after distance correction via a Bayesian approach (see Section~\ref{sec:dis_correct}). 
        The blue dots represent member stars in each cluster. The tidal radius of each cluster is indicated with a black circle.
        The dashed line indicates the direction of the line-of-sight. The grey dots in the background show the spatial distribution 
        of members without distance correction.
	    }
\label{fig:3d1}
\end{figure*}

\begin{figure*}[tb!]
\centering
\includegraphics[angle=0, width=\textwidth]{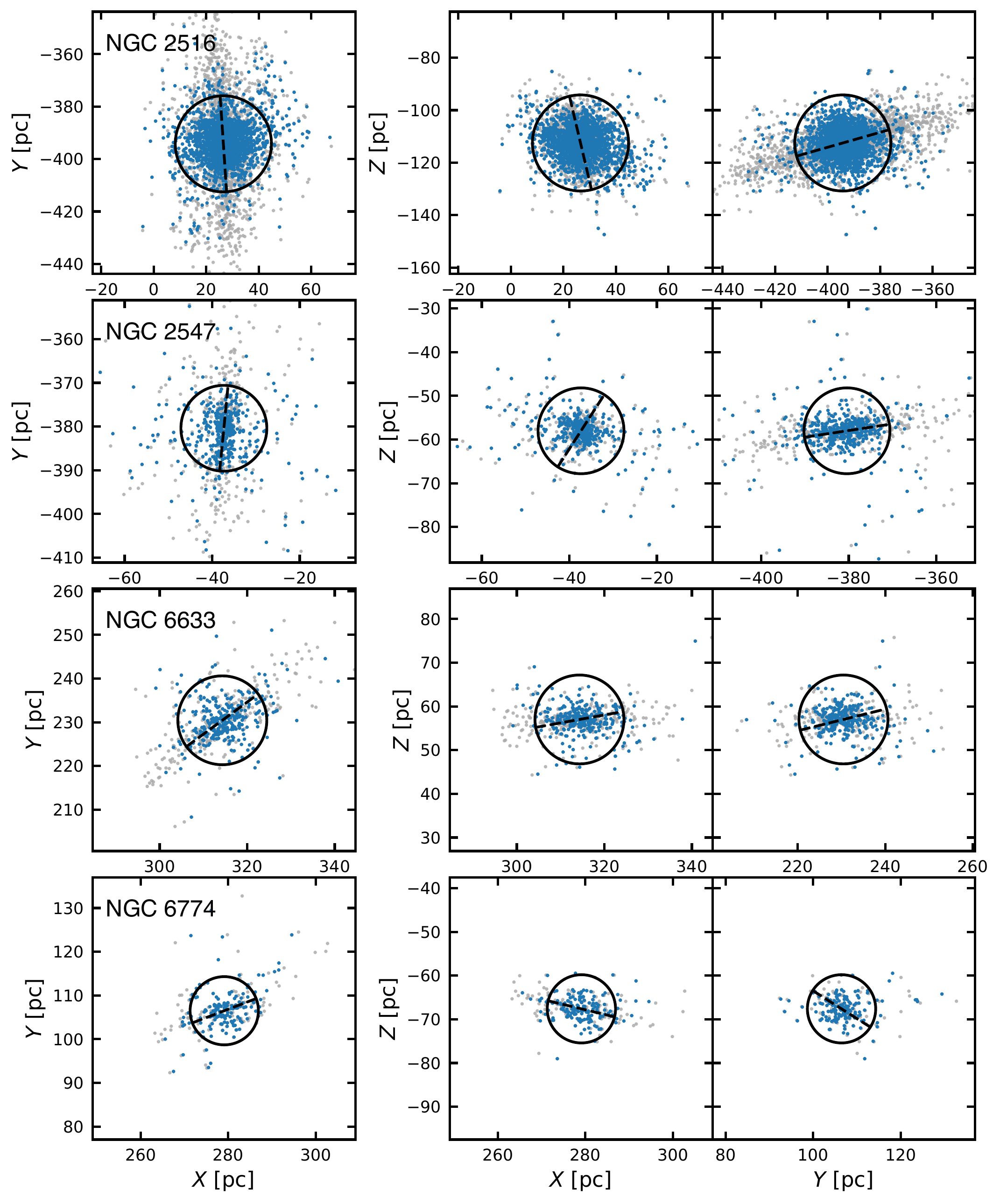}
\caption{3D spatial position of members in four target clusters: NGC\,2516 and NGC\,2547, NGC\,6633, NGC\,6774, 
        in heliocentric Cartesian coordinates ($X,Y,Z$; see definition in Appendix~\ref{sec:apx_coordinate}),
        after distance correction via a Bayesian approach (see Section~\ref{sec:dis_correct}). 
        Colors and symbols are the same as in Figure~\ref{fig:3d1}.  
        Filament-like substructures present in the young cluster NGC\,2547, and tidal-tail-like substructures in the 
        older clusters NGC\,2516, NGC\,6633, and NGC\,6774. }
\label{fig:3d2}
\end{figure*}

\begin{figure*}[tb!]
\centering
\includegraphics[angle=0, width=0.8\textwidth]{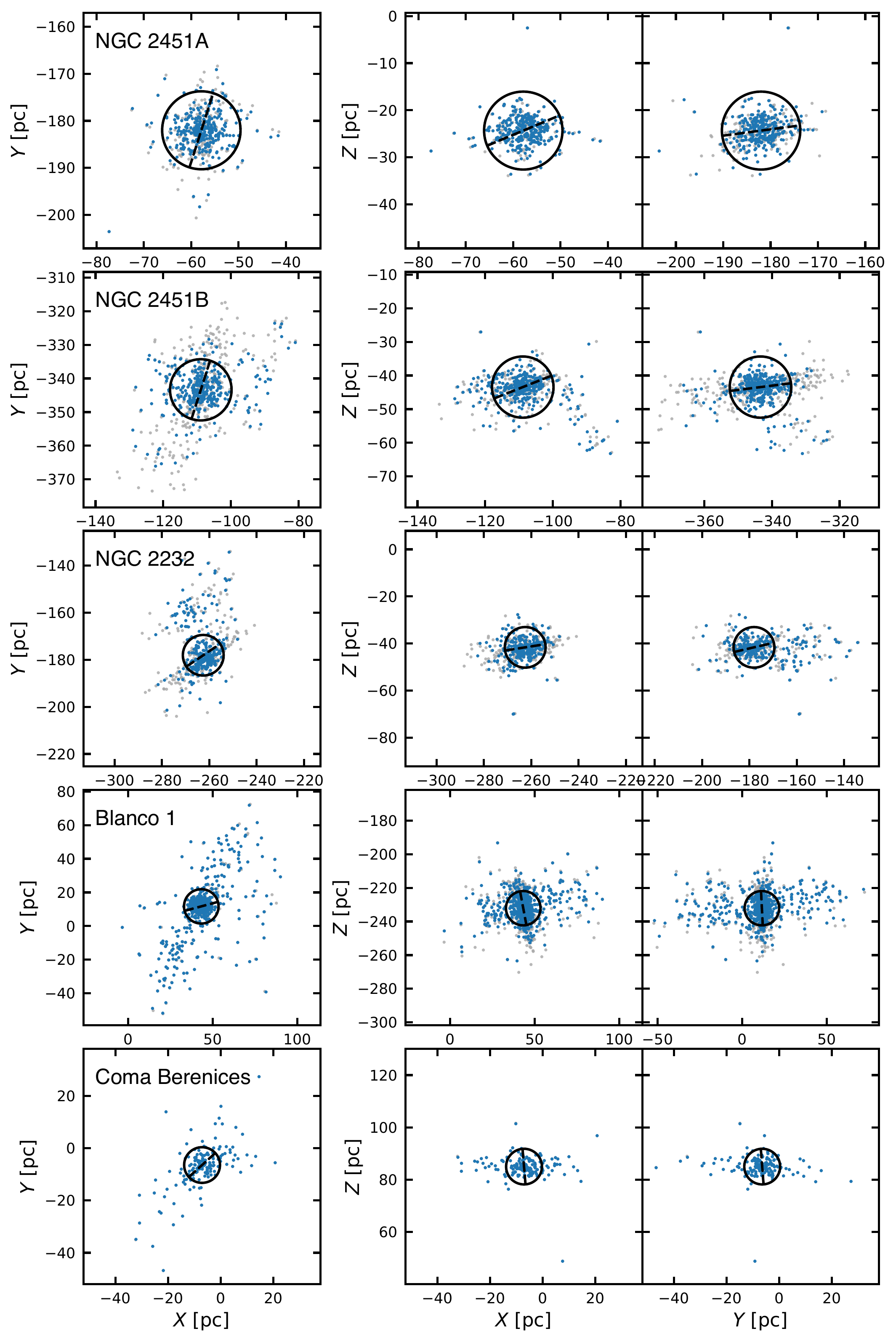}
\caption{3D spatial position of members in five target clusters: NGC\,2451A, NGC\,2451B, NGC\,2232, Blanco\,1 and Coma Berenices, 
        in heliocentric Cartesian coordinates ($X,Y,Z$; see definition in  Appendix~\ref{sec:apx_coordinate}), after distance 
        correction via a Bayesian approach (see Section~\ref{sec:dis_correct}). 
        Colors and symbols are the same as in Figure~\ref{fig:3d1}. Filament-like substructures present in the young clusters 
        NGC\,2451B and NGC\,2232, and tidal tails in the older clusters Blanco\,1 and Coma Berenices. 
	    }
\label{fig:3d3}
\end{figure*}

In Figures~\ref{fig:3d1}, \ref{fig:3d2} and \ref{fig:3d3} we show the 3D spatial distributions after correcting distances for member stars in the 13 target clusters. 
The corrected 3D positions of the members in all 13 target clusters are presented in Table~\ref{tab:memberlist}, together with other parameters from {\it Gaia} EDR\,3. We also present the 3D positions of the 13 target clusters in Table~\ref{tab:general_xyz_uvw}.  
The Bayesian method has provided a reasonable correction of the stretched shapes along the line-of-sight of each cluster (grey dots in Figures~\ref{fig:3d1},  \ref{fig:3d2} and \ref{fig:3d3}). 

To estimate the uncertainty in the corrected distance to each cluster, we carry out additional simulations  for each individual target cluster, with a uniform model (for details, see Appendix~\ref{sec:dis_correct_er}). We apply the mean parallax error to the members of each cluster and move the simulated cluster to the same distance as each target cluster. The corresponding uncertainty in the distance correction of each cluster (Table~\ref{tab:general}) is represented with a colored symbol in Figure~\ref{fig:simulation_dis_er}, that follows the curve of the uniform model. The most distant star cluster, NGC\,2422 (476\,pc), has an uncertainty of 3.2\,pc in the corrected distance. The uncertainty in the distance obtained through the Bayesian distance correction is much smaller than the error that arises from directly inverting {\it Gaia} parallax (see Section~\ref{sec:dis_correct}).

To quantify the size of each star cluster, we compute their tidal radii as
\begin{equation}
    r_t=\left( \frac{GM_{cl}}{2(A-B)^2}\right)^{\frac{1}{3}}\quad ,
\end{equation}
\citep{pinfield1998}. Here, $G$ is the gravitational constant, $M_{cl}$ is the total mass of the star cluster (i.e., the sum of the masses of the individual member stars),
and the parameters $A$ and $B$ are the Oort constants 
\citep[$A=15.3\pm0.4\rm~km~s^{-1}~kpc^{-1}$ and
$B=-11.9\pm0.4\rm~km~s^{-1}~kpc^{-1}$; see][]{bovy2017}. 

In the analysis below, we assume that candidate members located within tidal radius are gravitationally bound to the star cluster, while members outside are unbound. 
The mass of each individual member star is obtained from the nearest point in the fitted isochrone that is searched for using the $k$-D tree method \citep{millman2011}. The tidal radius of each cluster is indicated with a black circle in each panel of Figures~\ref{fig:3d1}, \ref{fig:3d2} and \ref{fig:3d3}.

%fig4

%fig5

%fig5

% Table 3
\begin{deluxetable*}{ccc}
\tablecaption{Columns for the table of corrected 3D positions of members in all target clusters. \label{tab:memberlist}
             }
\tabletypesize{\scriptsize}
\tablehead{
\colhead{Column}    & \colhead{Unit}    & \colhead{Description}
}
\startdata
Cluster Name                    &                  &  Name of the target cluster   \\
{\it Gaia} ID                   &                  &  Object ID in {\it Gaia} EDR\,3\\
ra                              & degree           &  R.A. at J2016.0 from {\it Gaia} EDR\,3\\
er\_RA                          & mas              &  Positional uncertainty in R.A. at J2016.0 from {\it Gaia} EDR\,3 \\
dec                             & degree           &  Decl. at J2016.0 from {\it Gaia} EDR\,3 \\
er\_DEC                         & mas              &  Positional uncertainty in decl. at J2016.0 from {\it Gaia} EDR\,3 \\
parallax                        & mas              &  Parallax from {\it Gaia} EDR\,3\\
er\_parallax                    & mas              &  Uncertainty in the parallax \\
pmra                            & mas~yr$^{-1}$    &  Proper motion with robust fit in $\alpha \cos\delta$ from {\it Gaia} EDR\,3     \\
er\_pmra                        & mas~yr$^{-1}$    &  Error of the proper motion with robust fit in $\alpha \cos\delta$   \\
pmdec                           & mas~yr$^{-1}$    &  Proper motion with robust fit in $\delta$ from {\it Gaia} EDR\,3     \\
er\_pmdec                       & mas~yr$^{-1}$    &  Error of the proper motion with robust fit in $\delta$  \\
Gmag                            & mag              & Magnitude in $G$ band from {\it Gaia} EDR\,3   \\
BR                              & mag              & Magnitude in $BR$ band from {\it Gaia} EDR\,3   \\
RP                              & mag              & Magnitude in $RP$ band from {\it Gaia} EDR\,3   \\
Gaia\_radial\_velocity          & km~s$^{-1}$      &  Radial velocity from {\it Gaia} DR\,2 \\
er\_Gaia\_radial\_velocity      & km~s$^{-1}$      &  Error of radial velocity from {\it Gaia} EDR\,3\\
Jackson\_radial\_velocity       & km~s$^{-1}$      &  Radial velocity from Gaia/ESO survey \citep{jackson2020}\\
er\_Jackson\_radial\_velocity   & km~s$^{-1}$      &  Error of radial velocity from Gaia/ESO survey \citep{jackson2020}\\
Bailey\_radial\_velocity        & km~s$^{-1}$      &  Radial velocity from \citet{bailey2018} \\
er\_Bailey\_radial\_velocity    & km~s$^{-1}$      &  Error of radial velocity from \citet{bailey2018} \\
Mass                            & M$_\odot$        & Stellar mass obtained in this study\\
X\_obs                          & pc               & Heliocentric Cartesian X coordinate computed via direct inverting {\it Gaia} EDR\,3 parallax $\varpi$ \\
Y\_obs                          & pc               & Heliocentric Cartesian Y coordinate computed via direct inverting {\it Gaia} EDR\,3 parallax $\varpi$ \\
Z\_obs                          & pc               & Heliocentric Cartesian Z coordinate computed via direct inverting {\it Gaia} EDR\,3 parallax $\varpi$ \\
X\_cor                          & pc               & Heliocentric Cartesian X coordinate after distance correction in this study \\
Y\_cor                          & pc               & Heliocentric Cartesian Y coordinate after distance correction in this study \\
Z\_cor                          & pc               & Heliocentric Cartesian Z coordinate after distance correction in this study \\
Dist\_cor                       & pc               & The corrected distance of individual member\\
\enddata
\tablecomments{A machine readable full version of this table is available online.}
\end{deluxetable*}

% Table 1
\begin{deluxetable*}{c RRR c RRR}%{r CLC c CLC}
\tablecaption{3D positions and velocities of 13 target clusters \label{tab:general_xyz_uvw}
             }
% \tablewidth{\dimen1000}
\tabletypesize{\scriptsize}
\tablehead{
	 \colhead{Cluster}  & 
	 \colhead{$X_m$}    & \colhead{$Y_m$}    & \colhead{$Z_m$}       &&
	 \colhead{$U$}      & \colhead{$V$}     & \colhead{$W$}     \\
%----------------
	 \colhead{}         &
	 \multicolumn{3}{c}{(pc)} & \colhead{}             &
	 \multicolumn{3}{c}{(km~s$^{-1}$)}                          \\
	 \cline{2-4} \cline{6-8} 
	 \colhead{(1)} & \colhead{(2)} & \colhead{(3)} & \colhead{(4)} && \colhead{(5)} & \colhead{(6)} & \colhead{(7)} 
	 }
\startdata
IC 2391	    &	1.22	&	-150.35	&	-17.92	&&	-23.88	&	-15.56	&	-5.75	\\
IC 2602	    &	50.61	&	-141.97	&	-12.98	&&	-7.92	&	-22.16	&	-0.73	\\
IC 4665	    &	283.76	&	167.91	&	101.81	&&	-3.58	&	-17.41	&	-8.85	\\
NGC 2422	&	-299.34	&	-369.47	&	26.22	&&	-30.66	&	-22.29	&	-10.86	\\
NGC 2516    &	26.65	&	-394.41	&	-112.85	&&	-21.99	&	-25.02	&	-4.51	\\
NGC 2547    &	-37.06	&	-381.18	&	-57.98	&&	-16.15	&	-9.99	&	-10.95	\\
NGC 6633    &	315.25	&	228.84	&	56.99	&&	-20.64	&	-17.66	&	-7.60	\\
NGC 6774    &	278.57	&	106.22	&	-67.20	&&	49.05	&	-19.42	&	-24.30	\\
NGC 2451A   &	-58.09	&	-182.22	&	-23.22	&&	-27.16	&	-14.30	&	-12.78	\\
NGC 2451B   &	-109.29	&	-343.86	&	-43.58	&&	-18.75	&	-8.28	&	-12.11	\\
NGC 2232    &	-261.38	&	-179.95	&	-41.41	&&	-20.44	&	-13.08	&	-10.86	\\
Blanco 1    &	42.91	&	11.44	&	-233.01	&&	-18.55	&	-6.62	&	-9.80	\\
Coma Berenices 	&	-7.25	&	-6.07	&	85.27	&&	-2.47	&	-5.63	&	-0.33	\\
\enddata
\tablecomments{$X_m$, $Y_m$, $Z_m$ is 3D position of 13 target clusters in the heliocentric Cartesian coordinates, taken as the median value of all members. $U$, $V$, $W$ are mean 3D velocities of each cluster in the heliocentric Cartesian coordinates.}
\end{deluxetable*}

The global morphology of an OC can generally be described with a dense central core (or nucleus) and an outer halo (or corona). The halo is much more extended and has a low stellar number density \citep{nilakshi2002}. However, the number of members in the halo can be substantial \citep{meingast2020}. Both Blanco\,1 and Coma Berenices show two grand tidal tails spanning up to 50--60\,pc from the cluster center, which belong to the halo region, accounting for more than 36\% and 50\% of their members, respectively. The direction of the tidal tails in Coma Berenices and Blanco\,1 are found to be parallel to the Galactic plane, in agreement with previous studies \citep{bergond2001, chen2004}. No apparent elongation is present in the young clusters IC\,2391, IC\,2602. IC\,2391 is more centrally compact showing a clear core, while IC\,2602 is more populous. Despite the age of 36\,Myr, IC\,4665 has a sparse distribution without a clear central concentration, which may be a consequence of rapid gas expulsion \citep[][see more discussion in Section~\ref{sec:nbody}]{pang2020, dinnbier2020a}. An elongated shape along the line-of-sight is apparent for the region containing the stars that are gravitationally bound to the cluster (i.e., inside tidal radius) for NGC\,2422, NGC\,2547, NGC\,6633, and Blanco\,1 (the angle between elongation and the line-of-sight, $\phi$, is presented in Table~\ref{tab:morph_kin}). The errors in the corrected distances to these clusters ($\approx$ 2.1--3.2 pc) are much smaller than the extent of their elongated regions ($\approx$20--30 pc). Therefore, the detection of the elongations are robust. Six clusters, IC\,2391, IC\,2602, NGC\,2457, NGC\,2451A, NGC\,2516 and Blanco\,1 overlap with previous work by \citet{meingast2020} based on {\it Gaia} DR\,2. Approximately 60--90\% of our members cross-match with members determined by \citet{meingast2020}. The majority of the matched members is located within tidal radius of the cluster. Our current member identification method is unable to confirm the membership of the stars in the vast extended stellar corona that were identified by \citet{meingast2020}. 

%around our young target clusters (IC\,2391, IC\,2602, NGC\,2457, NGC\,2451A). For example, the most massive cluster NGC\,2516 with 800 more members obtained in this work than in \citet{meingast2020}, its morphology resembles an oblate spheroid with a high density in the center, instead of an elongated shape extending $\sim$300\,pc from \citet{meingast2020}'s study.

 As a result of the higher accuracy of the proper motion measurements in {\it Gaia} EDR\,3,  the extended filamentary structures of NGC\,2232 that were once identified as two separate groups (purple and green) in \citet{pang2020} (using {\it Gaia} DR\,2 data and with the same selection technique) are now identified as members of NGC\,2232. This confirms the conclusion of \citet{pang2020} that the coeval filamentary structures are closely related to NGC\,2232, which are formed at the same time in the parental molecular clouds \citep{jerabkova2019,beccari2020,tian2020}. Similar filament-like substructures are also found in another two young clusters, NGC\,2547 and NGC\,2451B. On the other hand, tidal-tail-like structures extending up to 10--20\,pc are detected in three older clusters: NGC\,2516, NGC\,6633 and NGC\,6774. 
The diffuse spatial distribution of the oldest cluster (NGC\,6774) implies its advanced dissolution state, after having experienced substantial secular dynamical evolution. The 3D morphology of OCs again confirms the presence of the 2D elongation that we observed in NGC\,2547 and NGC\,2516, NGC\,2232 and NGC\,2451B in Figure~\ref{fig:lb}. 
 
%Similar work has been carried out by \citet{meingast2020} that identify members and correct for distance error along the line-of-sight with a Gaussian mixture method with {\it Gaia} DR\,2. We introduce a pioneering member-identification method that is informed by cluster bulk velocities and deconvolves the spatial distribution with a mixture of Gaussians. Our approach enables inferring the true spatial distribution of the clusters by effectively filtering field star contaminants while at the same time mitigating the effect of positional errors along the line of sight.

%-------------------------------------------------------------------------------------------%
%-------------------------------------------------------------------------------------------%
\subsection{Parameterization of 3D morphologies}\label{sec:par_3d}

\begin{figure*}[tb!]
\centering
\includegraphics[angle=0, width=0.75\textwidth]{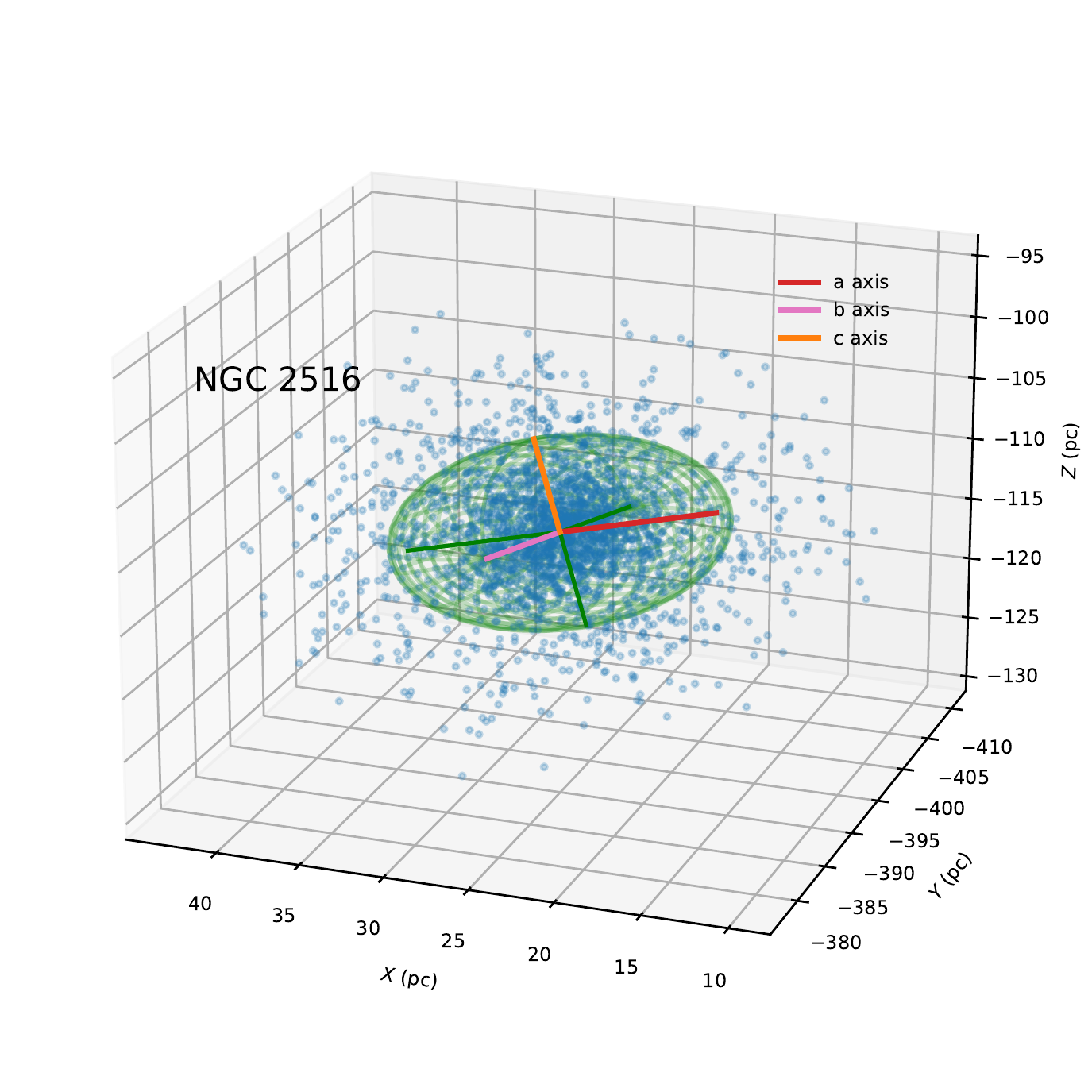}
\caption{
        Ellipsoid fitting for the 3D spatial positions in heliocentric Cartesian cooridinates, ($X,Y,Z$), for the cluster members within tidal radius of NGC\,2516, after distance correction through a Bayesian approach (see Section~\ref{sec:dis_correct}). The green surface represents the fitted ellipsoid. Blue dots are members within tidal radius. The three axes of the ellipsoid ($a$, $b$, and $c$) are indicated in red, pink, and orange, respectively. 
	    }
\label{fig:ellipsoid_tidal}
\end{figure*}

%-------------------------------------------------------------------------------------------%
From the 3D distribution of member stars in each cluster (Figures~\ref{fig:3d1}, \ref{fig:3d2}, and \ref{fig:3d3}), the general shape of member distribution within the tidal radius can be approximated with an ellipsoid. We perform ellipsoid fitting\footnote{\url{https://github.com/marksemple/pyEllipsoid_Fit}} to the 3D morphology of each cluster in order to quantify the shape of the distribution of bound stars in the target clusters (we do not include the members located outside tidal radius, since their number is small). NGC\,2516 is shown as an example to illustrate the ellipsoid fitting for the bound member stars  (see Figure~\ref{fig:ellipsoid_tidal}). The fitted ellipsoid (green surface) is centered at the median position of bound members, which we consider as the cluster center.  The three semi-axes of the ellipsoid $a$, $b$, $c$ are the free parameters in this fit, where $a$ is the semi-major axis (red line), $b$ the semi-intermediate axis (pink line), and $c$ the semi-minor axis (orange line). We use the lengths of the semi-axes $a$, $b$, $c$, and axis ratios $b/a$ and $c/a$ to describe the morphology of the clusters, and the direction of the semi-major axis $a$ of the fitted ellipsoid as the direction of elongation of the star cluster. Smaller values of the axis ratios $b/a$ and $c/a$ indicate a more elongated structure. The fitted values of the morphological parameters ($a$, $b$, $c$, $b/a$ and $c/a$) are listed in Table~\ref{tab:morph_kin}. We also compute for each cluster the angle $\theta$ between the direction of $a$ and the Galactic plane (the projection of $a$ on the Galactic plane), and the angle $\phi$ between the direction of $a$ and the line-of-sight. The values of these two angles are listed in Table~\ref{tab:morph_kin}. The fitted ellipsoids for stars inside the tidal radius of other twelve target OCs are presented in Appendix~\ref{sec:apx_ellipsoid} (Figure~\ref{fig:apx_ellip_all}).

 %fig4

% Table 2
\begin{deluxetable*}{cRRR c RR c RR c LLL}
\tablecaption{Morphological and kinematic parameters of target clusters \label{tab:morph_kin}
             }
% \tablewidth{\dimen1000}
\tabletypesize{\scriptsize}
\tablehead{
	 \colhead{Cluster Name} & \colhead{$a$}     & \colhead{$b$} & \colhead{$c$} & 
	 \colhead{}             &
	 \colhead{$b/a$}        & \colhead{$c/a$}   & 
	 \colhead{}             &
	 \colhead{$\theta$}     & \colhead{$\phi$}  & 
	 \colhead{}             & 
	 \colhead{$\sigma_{RV}$}& \colhead{$\sigma_{pmra}$} & \colhead{$\sigma_{pmdec}$}        \\
	 %------------------
	 \colhead{}             & \multicolumn{3}{c}{(pc)} &
     \colhead{}             &
	 \multicolumn{2}{c}{(axis ratio)}                  & 
	 \colhead{}             & 
	 \multicolumn{2}{c}{(degrees)}                     & 
	 \colhead{}             & 
	 \multicolumn{3}{c}{(km~s$^{-1}$)}                                                      \\
	 \cline{2-4} \cline{6-7} \cline{9-10} \cline{12-14} 
	 \colhead{(1)} & \colhead{(2)} & \colhead{(3)} &
	 \colhead{(4)} && \colhead{(5)} &
	 \colhead{(6)} && \colhead{(7)} &
     \colhead{(8)} && \colhead{(9)} & \colhead{(10)} & \colhead{(11)} 
	 }
\startdata
IC 2391	    &	5.43\pm0.16	&	3.49\pm0.06	&	2.21\pm0.05	&&	0.64\pm0.02	&	0.41\pm0.02 && 16.76    & 46.94 &&  0.34^{+0.22}_{-0.20}    & 0.50^{+0.07}_{-0.06}  & 0.43\pm0.06           \\
IC 2602	    &	5.48\pm0.11	&	4.37\pm0.04	&	3.66\pm0.05	&&	0.80\pm0.02	&	0.67\pm0.02 &&  5.44    & 41.35 &&  0.20^{+0.15}_{-0.13}    & 0.50^{+0.07}_{-0.06}  & 0.50^{+0.07}_{-0.06}  \\
IC 4665	    &	6.13\pm3.03	&	4.49\pm0.40	&	3.72\pm0.48	&&  0.73\pm0.37	&	0.61\pm0.31	&& 38.78    & 22.50 &&  0.38\pm0.16             & 0.40^{+0.07}_{-0.05}  & 0.28\pm0.05           \\
NGC 2422	&	6.30\pm4.04	&	5.54\pm0.49	&	4.57\pm0.56	&&  0.88\pm0.57	&	0.73\pm0.47	&& 19.01    & 79.28 &&  0.71^{+0.18}_{-0.15}    & 0.47^{+0.07}_{-0.05}  & 0.50\pm0.07           \\
NGC 2516	&	10.03\pm3.75&	9.32\pm0.48	&	6.98\pm0.53	&&	0.93\pm0.35	&	0.70\pm0.27	&& 11.93    & 31.29 &&  0.72\pm0.07             & 0.82\pm0.04           & 0.80\pm0.04           \\
NGC 2547	&	7.42\pm2.78	&	6.14\pm0.39	&	2.88\pm0.44	&&  0.83\pm0.31	&	0.39\pm0.16	&&  4.20    & 10.81 &&  0.68^{+0.09}_{-0.08}    & 0.39\pm0.04           & 0.42\pm0.04           \\
NGC 6633	&	8.20\pm2.21	&	5.60\pm0.31	&	3.34\pm0.38	&&  0.68\pm0.19	&	0.41\pm0.12	&&  1.22    & 23.90 &&  0.76^{+0.32}_{-0.22}	& 0.47^{+0.09}_{-0.07}  & 0.77\pm0.19           \\
NGC 6774	&	6.31\pm1.45	&	4.37\pm0.22	&	2.98\pm0.27	&&  0.69\pm0.16	&	0.47\pm0.12	&& 18.30    & 17.53 &&  0.55^{+0.17}_{-0.15}    & 0.39\pm0.04           & 0.71^{+0.09}_{-0.07}  \\
NGC 2451A	&	5.43\pm0.56	&	4.98\pm0.11	&	3.52\pm0.12	&&  0.92\pm0.10	&	0.65\pm0.07	&& 14.33    & 81.06 &&  0.23^{+0.35}_{-0.17}    & 0.68^{+0.10}_{-0.08}  & 0.37^{+0.06}_{-0.05}  \\
NGC 2451B	&	5.96\pm2.62	&	5.69\pm0.37	&	4.14\pm0.42	&&  0.95\pm0.42	&	0.69\pm0.31	&& 22.12    & 48.56 &&  0.25^{+0.20}_{-0.16}    & 0.48^{+0.07}_{-0.05}  & 0.34^{+0.05}_{-0.03}  \\
NGC 2232	&	6.11\pm2.01	&	5.12\pm0.28	&	3.37\pm0.36	&&  0.84\pm0.28	&	0.55\pm0.19	&&  4.55    & 26.05 &&  0.10^{+0.11}_{-0.07}	& 0.27^{+0.05}_{-0.03}  & 0.29^{+0.05}_{-0.03}  \\
Blanco~1	&	8.28\pm1.66	&	4.65\pm0.25	&	4.01\pm0.31	&&  0.56\pm0.12	&	0.48\pm0.10	&& 78.35    & 12.53 &&  0.32\pm0.08     	    & 0.40\pm0.03           & 0.36\pm0.03           \\
Coma Berenices	&	4.91\pm0.02	&	4.23\pm0.02	&	3.43\pm0.01	&&	0.86\pm0.01	&	0.70\pm0.01	&& 14.3 & 79.38 &&  0.61\pm0.18	            & 0.22^{+0.04}_{-0.03}  & 0.32^{+0.05}_{-0.04}  \\
\enddata
\tablecomments{$a$, $b$, $c$ are the semi-major, semi-intermediate and semi-minor axes of the fitted ellipsoid for each star cluster in the sample. $\theta$ is the angle between the direction of $a$ and the Galactic plane. The quantity $\phi$ is the angle between the direction of $a$ and the line-of-sight. $\sigma_{RV}$ is the RV dispersion (within the tidal radius);  $\sigma_{pmra}$ and $\sigma_{pmdec}$ are the dispersions of the R.A. and Decl. components of the PMs (within the tidal radius). The values in columns 9--11 are obtained using the MCMC method; each best-fit value is the median of the posterior distribution, and the uncertainties are the corresponding 16- and 84-percentiles of the posterior.}
\end{deluxetable*}

The clusters NGC\,2516, NGC\,2547, NGC\,2451A, NGC\,2451B and NGC\,2232 have axes ratios of approximately $b/a=0.8-0.95$, while $c/a=0.4-0.7$. The morphologies of these five clusters resemble oblate  spheroids. The other clusters, IC\,2602, IC\,4665, and NGC\,2422, have shapes that are well described by prolate spheroids, with a difference between $b/a$ and $c/a$ of less than $\sim$10\%. After excluding the prominent tails outside the tidal radius for Coma Berenices and Blanco\,1, prolate spheroidal distributions fit both clusters. The morphologies of the remaining clusters (IC\,2391, NGC\,6633 and NGC\,6774) can be approximated as triaxial ellipsoids.  

The non-spherical shape for the bound region of most target clusters is likely a result of the interplay of internal and external dynamical processes. Relaxed stars gradually evaporate, primarily through the Lagrange points \citep{kupper2008}, a process that depends on the motion of the each cluster through the Galactic disk. In addition, the external tidal field exerts a force that pulls the cluster apart along the axis that connects the cluster to the Galactic center. Due to differential rotation, the induced tidal tails always tilt with respect to the cluster orbit \citep{tang2019}. That is the reason why the tidal tails of Coma Berenices and Blanco\,1 are parallel to the Galactic plane (see  Figure~\ref{fig:3d3}). The bound region (within the tidal radius) of most clusters has an elongation direction that is more or less aligned with the Galactic plane (see Figure~\ref{fig:apx_ellip_all}), with an angle (between $a$ and the disk) of $|\theta|<20^\circ$ (see Table~\ref{tab:morph_kin}). This result is thus in agreement with earlier findings \citep{oort1979, bergond2001, chen2004}, and also indicates that despite their young age, most clusters have already been affected by the external Galactic tides. The direction of the semi-major axis $a$ does not align with the direction of the line-of-sight (see the values of $\phi$ in Table~\ref{tab:morph_kin}), confirming the reliability of our distance correction (Section~\ref{sec:dis_correct}).

Although Blanco\,1 appears to show evidence of having been affected by the tidal force, its bound region has an elongation that is closely aligned with the vertical ($Z$) direction of perpendicular to the Galactic plane  (with an angle of $\approx11.7^\circ$). While stars escape mainly through the two Lagrange points, the evaporation process stretches all the way between the two Lagrange points \citep{kupper2008} and generates the elongated shape in the distribution of the bound stars. The unbound stars are subjected to Galactic tides so that their orbits become more tangential and form the tidal tails around Blanco\,1, which probably constitutes both ``tail I'' and ``tail II'' \citep{dinnbier2020a}.

The distribution of the bound population of stars in the oldest cluster NGC\,6774 can be describe with an  triaxial ellipsoid. The secular relaxation process in NGC\,6774 results in significant mass loss, that results in the formation of tidal-tail-like structures beyond the tidal radius \citep[see Figure~\ref{fig:3d2} and][]{yeh2019}, and the escape velocity is consequently greatly reduced. A phase of global evaporation must have taken place.

%%%%%%%%%%%%%%%%%%%

Internal stellar dynamics, such as two-body relaxation, tend to produce an isotropic velocity distribution in the radial direction.  Therefore, the core of the OC becomes more spherical as it evolves. \citet{chen2004} indeed noted that the projected flattening of OCs decreases as cluster grow older. The axis ratios $b/a$ and $c/a$ provide appropriate tools to investigate this phenomenon, since they probe the bound region of OCs where internal dynamical processes dominate. However, among the 13 target clusters, no correlation appears to exist between $b/a$ and age. 
One may expect a decreasing trend between elongation and age if star clusters inherit their elongated shape from their parent GMC.
%When elongation is more pronounced in young OCs, this supports the scenario that young clusters inherit the elongated shape from their parental GMC.
As OCs evolve toward older age, their initial shape is ``forgotten'' as internal relaxation processes increase the sphericity of the clusters, especially the shape of the bound region. This process continues until the time when evaporation becomes the dominant process in the evolution of the cluster. 

A larger sample of OCs is required to further quantify the relation between morphology and cluster dynamics. Additional work is also needed to further deepen our understanding of the evolution of the shapes of OCs as they reside in the Galactic field. Numerical modelling will be a useful tool to obtain a theoretical benchmark in the future.

%-------------------------------------------------------------------------------------------%
\section{Dynamical states of open clusters}\label{sec:dynamics}

\subsection{3D velocity and velocity dispersion}\label{sec:dispersion}

The observed 3D morphology of OCs is thought to be driven by cluster dynamics. However, few studies have been carried out to investigate the relationship between cluster morphology and stellar dynamics. In this study, we connect the 3D morphology and the dynamical state of open clusters for the first time.
We use PMs and RVs from {\it Gaia} EDR\,3, and RVs from the literature, as discussed in Section~\ref{sec:gaia_member}.

\begin{figure*}[p]
\centering
\includegraphics[angle=0, width=1.\textwidth]{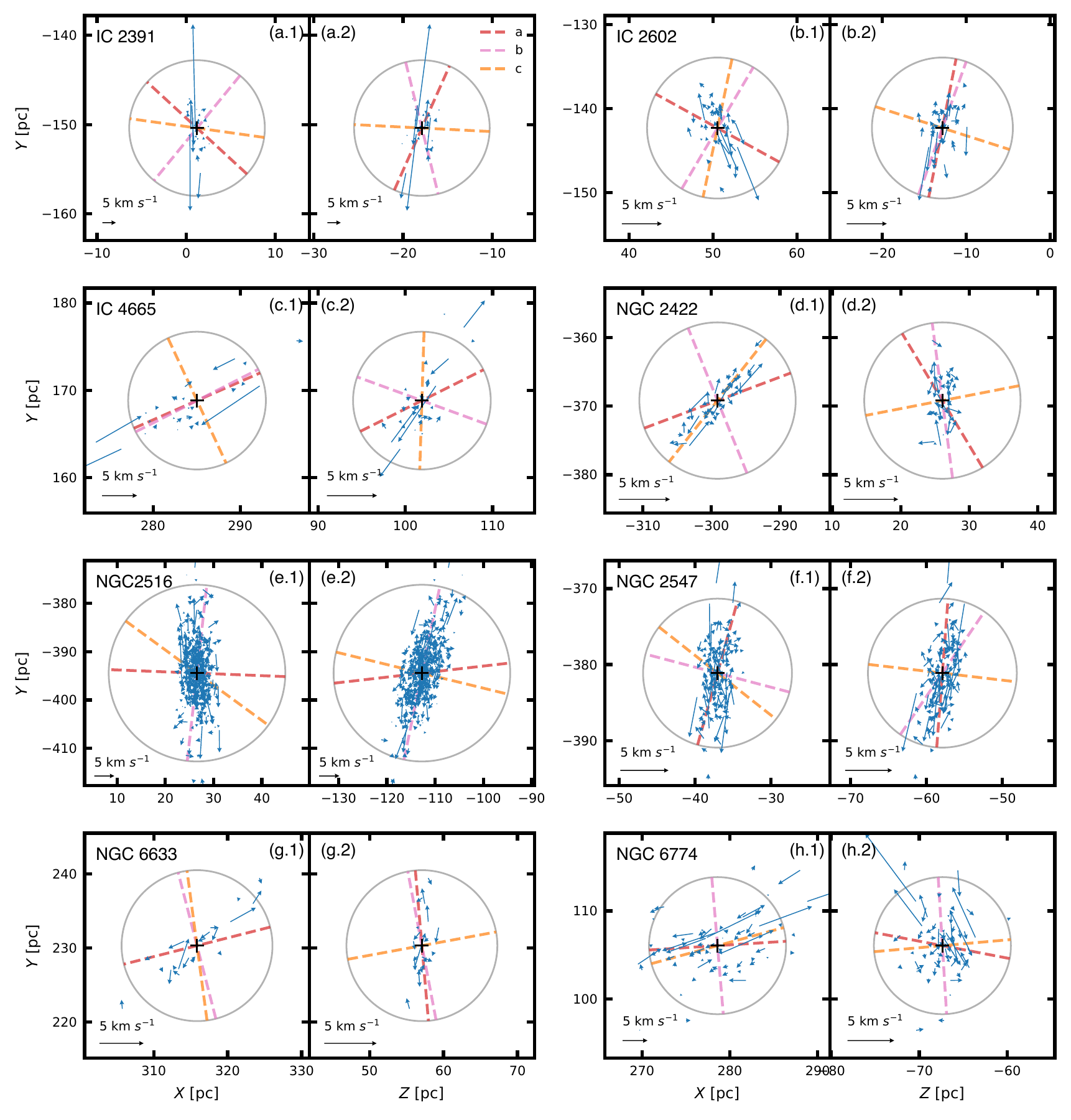} %vectors1_new.pdf
\caption{
        The relative 3D velocity vectors for members of eight target clusters, projected onto $X$-$Y$ and $Y$-$Z$ planes.
	    The blue vectors represent the velocities of member stars, relative to the mean motion of each cluster. The center of each cluster is indicated with the (+) symbol. Black circles denote the tidal radii. The scale of the velocity vectors is indicated in the bottom-left corner of each panel.	     
	    }
\label{fig:3d_vel1}
\end{figure*}

\begin{figure*}[tb!]
\centering
\includegraphics[angle=0, width=1.\textwidth]{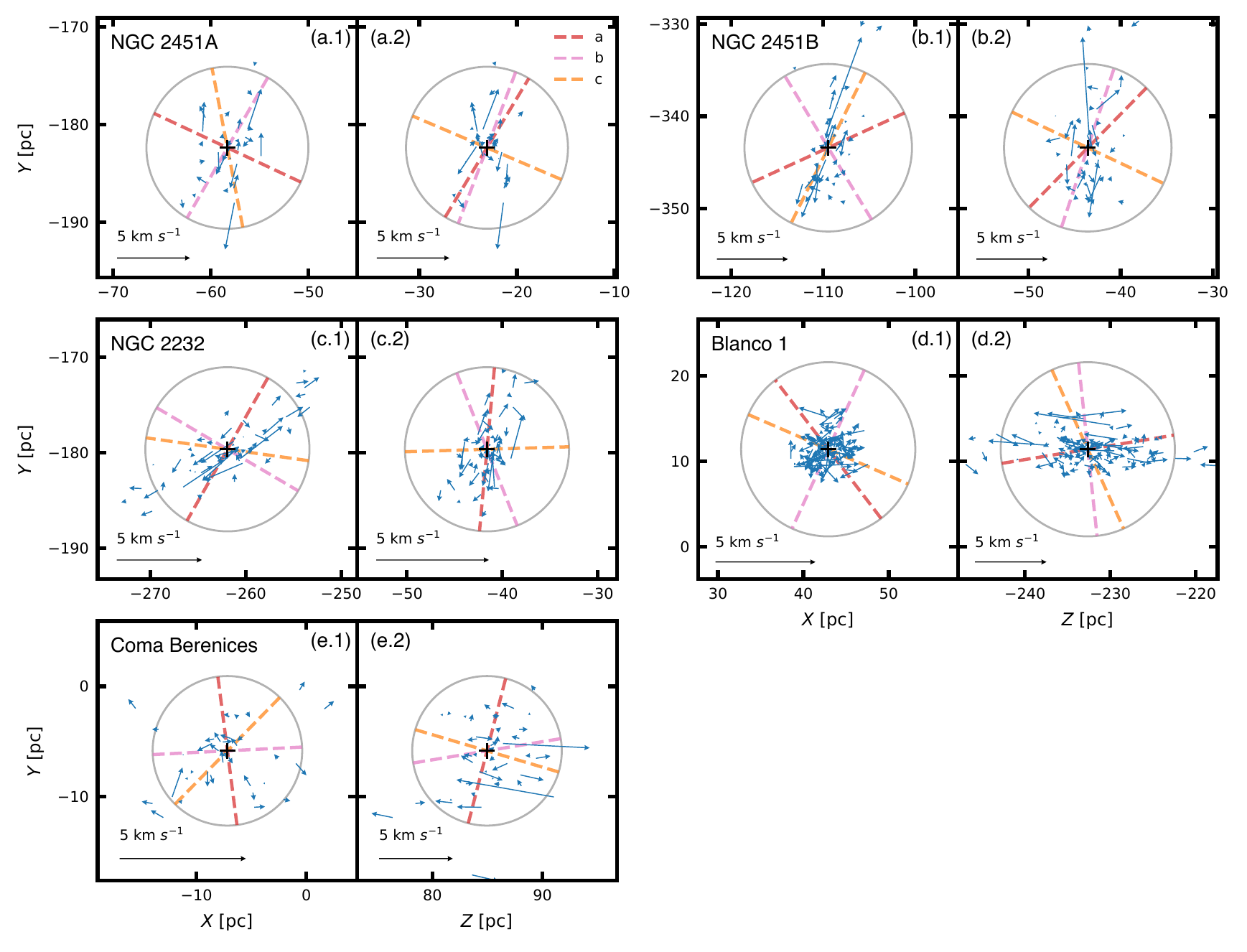}
\caption{
        The relative 3D velocity vectors for members of five target clusters, projected onto $X$-$Y$ and $Y$-$Z$ planes.
	    Colors and symbols are identical to those in Figure~\ref{fig:3d_vel1}	     
	    }
\label{fig:3d_vel2}
\end{figure*}

Considering extended structures in target clusters, we adopt the median position of the members in each cluster as the cluster centers, and we use the average velocity in each cluster as the origin of the reference frame,  values of which are listed in Table~\ref{tab:general_xyz_uvw}. We present the 3D velocity vectors of the member stars superposed on the 3D spatial positions (relative to that of the cluster center) in Figures~\ref{fig:3d_vel1} and~\ref{fig:3d_vel2}. The tidal radii and the projections of the $a$, $b$ and $c$ axes of the fitted ellipsoids of each cluster are overplotted. The majority of the members with 3D velocity measurements are located within the tidal radius. Members with 3D velocities that differ more than 2$\sigma$ from the mean value are excluded from the velocity vector plots. These high-velocity stars all have large RVs. They are most likely binary candidates \citep[see, e.g.,][for details]{kouwenhoven2008}, and are also located on the binary sequence in the CMD. One star with extraordinary RV is a blue straggler candidate in NGC\,6774. Its peculiar velocity may have originated from a close encounter or from a merger event.

Figures~\ref{fig:3d_vel1} and~\ref{fig:3d_vel2} show that the directions of the velocity vectors of a large number of members align with the major axis of the fitted ellipsoid that coincides with the direction of elongation. All clusters from young to old, are expanding, as the majority of members are seen to move away from the cluster center. Expansion in young clusters is thought to be driven by gas expulsion \citep{baumgardt2007, dinnbier2020a, dinnbier2020b, pang2020}. After the gas expulsion, member stars expand radially and therefore reduce the depth of the gravitational potential wells of the OCs. 

%fig4

%fig4

In order to quantitatively analyse the dynamical states of our target OCs, we compute the RVs and PMs' dispersion of the bound members in each cluster. 
The likelihood function for the RV distribution is a combination of two Gaussian components, one for cluster members, and the other one for field stars \citep[equations 1 and 8 in][]{cottaar2012}. The Gaussian distribution of cluster members is broadened by the orbital motions of unresolved binary systems, and also by the uncertainties in the RV measurement. To model the broadening introduced by binary stars, we adopt distributions of orbital parameters that are characteristic for solar-type stars in the Galactic field: (i) a log-normal orbital period distribution for the binaries \citep{raghavan2010}; (ii) a flat mass ratio distribution between $q=0$ and $q=1$ \citep{duchene2013}; and (iii) a flat eccentricity distribution between $e=0$ and the maximum value \citep{parker2009}. The adopted parameters of binary stars are more computationally efficient but still comparable to the more realistic models \citep[e.g.,][]{marks2011b,marks2011}. However, as pointed out by \citet{bravi2018}, the selected binary properties do not significantly affect the final fitted results. The likelihood function of the PM distribution ($\mu_\alpha \cos\delta$ and $\mu_\delta$) is described by two Gaussian profiles \citep[Equations 1--3 in][]{pang2018}: the component of the cluster members, and the field component (the latter accounts for 5\%). We use the Markov Chain Monte Carlo (MCMC) method to obtain the best-fit values and the corresponding uncertainties for the RV and PM velocity dispersions (columns 9--11 in Table~\ref{tab:morph_kin}).

The derived velocity dispersions can be used to quantify the rate of expansion of each cluster. Clusters with a higher velocity dispersion tend to expand faster. Based on the half-mass radius, $r_{\rm h}$ (Table~\ref{tab:general}), and the 3D velocity dispersion of each cluster (Table~\ref{tab:morph_kin}), we estimate their dynamical masses using Equation~1 in \citet{fleck2006}. The resulting dynamical masses (M$_{\rm dyn}$) of each cluster ranges from 263\,M$_\sun$ to 3368\,M$_\sun$ (Table~\ref{tab:general} column~8), higher than the estimated photometric masses, 101\,M$_\sun$ to 1973\,M$_\sun$ (column~7 in Table~\ref{tab:general}). This discrepancy is not resolved, even when we correct the mass of faint stars below {\it Gaia} EDR\,3's detection limit by extrapolating the mass function \citep[see demonstration in][]{tang2019}. Therefore, this suggest the majority of clusters might be supervirial and may end up expanding as the kinetic energy overtakes the gravitational potential. 

\begin{figure}[tb!]
\centering
\includegraphics[angle=0, width=1.\columnwidth]{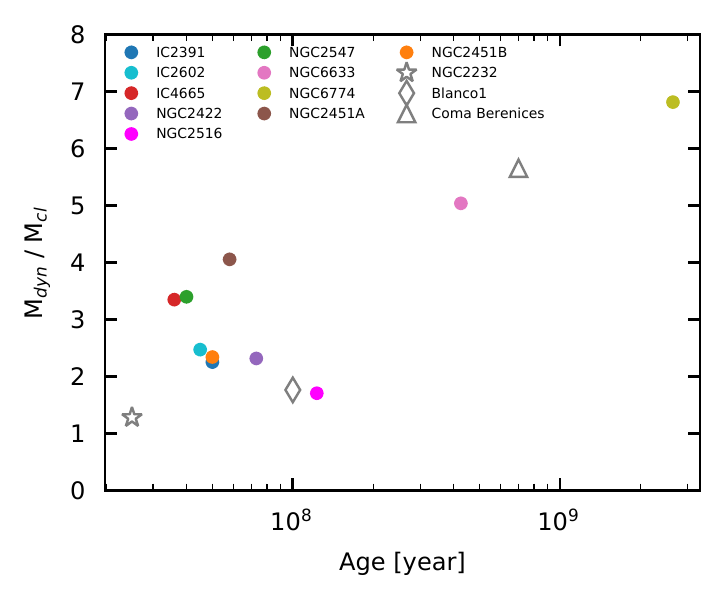}
\caption{ The relation between the ratio of dynamical mass over photometric mass, 
            M$_{\rm dyn}$/M$_{\rm cl}$, and cluster age. The large values of the ratio 
            suggests that most of the clusters may be super-virial.
}
\label{fig:mass_ratio}
\end{figure}

We display the dependence of ratio between the dynamical mass and the  photometric mass on the cluster age in Figure~\ref{fig:mass_ratio}. The ratio M$_{\rm dyn}$/M$_{\rm cl}$ increases as clusters grow older, especially after 300\,Myr. The oldest cluster in the sample, NGC\,6774, has the highest ratio of M$_{\rm dyn}$/M$_{\rm cl}$, further confirms its state of disruption. On the contrary, the youngest cluster in the sample, NGC\,2232,  has the lowest ratio (M$_{\rm dyn}$/M$_{\rm cl}\sim1.2$), which is consistent with the scenario suggested by \citet{pang2020} that this cluster is likely undergoing a phase of revirialization. The cluster probably reaches its maximal expansion before its re-collapse to form a virialised cluster  \citep[e.g.,][]{kroupa2001}.

%fig4
 
As suggested from simulations \citep{baumgardt2007}, stellar members will acquire highly-anisotropic velocity dispersions after rapid gas expulsion, and isotropic velocity dispersion after slow gas removal. Some degree of velocity anisotropy is observed in target clusters with detected elongated structures (NGC\,2547, NGC\,2451B, NGC\,2516, NGC\,6633, NGC\,6774, Blanco\,1, Coma Berenices) but not in NGC\,2232 (Table~\ref{tab:morph_kin}, columns 9--11). Velocity anisotropy may also originate from global rotation in star clusters. Global rotation is not uncommon; it was recently discovered in the open cluster Tr\,15 \citep{kuhn2019}. Although OCs may inherit angular momentum from parental GMCs, the merging substructures, or merging clusters \citep[e.g.,][]{priyatikanto2016, zhong2019, darma2019}, few attempts have been made to measure rotation in OCs. Unlike OCs, rotation has been measured in globular clusters \citep[e.g.,][]{bianchini2018,kamann2018}.  
According to $N$-body simulations by \citet{einsel1999} and \citet{hong2013}, rotation enhances mass loss and therefore speeds up the disruption process of clusters. The  global rotation speeds are typically much lower in OCs than globular clusters, down to sub~km~s$^{-1}$ level. Higher-resolution of spectroscopy is required in order to quantify the rotational properties of our target OCs.

\subsection{Comparison with numerical models}\label{sec:nbody}

In order to determine the properties of the gas expulsion process of the 13 target clusters in this study, we carry out $N$-body simulations of OCs with four different sets of initial conditions, and compare our numerical findings with the observations.

\subsubsection{Initial conditions}

The initial mass $M_{\rm cl}(0)$ of the models is chosen such that the cluster mass $M_{\rm cl}$ at an evolved stage at the age $t$ is comparable to the mass of the observed clusters. Accordingly, we adopt $M_{\rm cl}(0) = 250\,{\rm M_\odot}$, $500\,{\rm M_\odot}$, $1000\,{\rm M_\odot}$, $2000\,{\rm M_\odot}$, and $4000\,{\rm M_\odot}$, which is consistent with the simulations of cluster-formation in molecular clouds \citep{bate2012}. All the cluster models are initialized with a Plummer model in virial equilibrium \citep{Aarseth1974a}, that is characterised by the initial cluster mass $M_{\rm cl}(0)$ and the half-mass radius $r_{\rm h}$. We note that a much larger variety of initial conditions, including non-spherically symmetric substructures, are possible \citep[e.g.][]{Moeckel2010,Fujii2015a}. However, such the substructure typically disappears quickly \citep[e.g., through feedback from photoionization;][]{gonzalez2020}, as the cluster relaxes and obtains a spherically symmetric configuration \citep{kroupa2001,Goodwin2004,Sills2018,Banerjee2018}. This process occurs prior to the onset of gas expulsion in our models. Since the uncertainty of the mechanism expelling the gas has likely a much more prominent impact on the cluster dynamics than the initial substructure, we focus on the gas expulsion mechanism in spherical systems in the present work.

Stellar masses are sampled from the \citet{Kroupa2001a} initial mass function (IMF), with a minimum mass of $m_{\rm min}=0.08\,{\rm M}_\odot$, and a maximum mass that is obtained following the $m_{\rm max} - M_{\rm cl}$ relation of \citet{Weidner2013}, where $m_{\rm max}$ is the maximum mass of a star formed in a cluster of mass $M_{\rm cl}$. We assume a binary fraction of 100\% among member stars \citep[see, e.g.,][]{goodwin2005} and initial distributions for the orbital elements (periods or binding energies, mass-ratios, and  eccentricities) derived from unifying the observed populations in very young populations that are in different stages of dynamical processing with the Galactic field population \citep{kroupa1995a, kroupa1995b, kroupa2001, marks2011, belloni2017}. 

The clusters move on circular orbits through the Galaxy at galactocentric radius $d_{\rm GC} = 8$\,kpc, and with an orbital speed of 220~km~s$^{-1}$. The simulated clusters are evolved until $t=100$\,Myr; dynamical evolution is most prominent during this time span. This age range covers two thirds of the age of our target clusters. All the present models are initialized as embedded star clusters, i.e. containing both the stellar and gaseous components. Over time, the gas is removed from the cluster due to feedback from massive stars. Technical details of the $N$-body simulations are described in Appendix~\ref{sec:apx_nbody}. 

In the first model~S0, the initial half-mass radius $r_{\rm h}$ of the cluster is related to the initial cluster mass $M_{\rm cl}(0)$, following the relation of \citet{Marks2012}. These initial conditions generate star clusters of rather compact sizes ($r_{\rm h}\approx 0.3$\,pc) for clusters with initial mass $M_{\rm cl}(0) = 4000\,{\rm M_\odot}$. 
Model~S0 has a star formation efficiency (SFE) of $1/3$ and a gas expulsion timescale $\tau_{\rm M} = 0.03$\,Myr, which removes the gas on a time-scale shorter than the stellar crossing time. In other words, the 
gas expulsion is impulsive. No primordial mass segregation is present in this model. 

The clusters in the second model~S5 are identical to model~S0, apart from its primordial mass segregation.
Mass segregation is generated using the method of \citet{Subr2008}, with an initial mass segregation index of $S = 0.5$. 
The clusters in the third scenario (model~AD) have a longer gas expulsion time-scale of $\tau_{\rm M} = 1$\,Myr. Thus, the gas is removed on a time-scale longer than stellar crossing time; i.e., the gas expulsion is adiabatic. 
Adiabatic gas expulsion typically impacts the cluster less than impulsive gas expulsion of the same SFE \citep[e.g.,][]{baumgardt2007,dinnbier2020b}. 
Clusters in the fourth scenario (model WG) contain no gas, i.e. $M_{\rm gas}(0) = 0$ and $\mathrm{SFE} = 0$, 
and have a large initial $r_{\rm h}$ of $1$\,pc.

\subsubsection{Comparison with target clusters}

\begin{figure*}[p]
\centering
\includegraphics[angle=0, width=1.\textwidth]{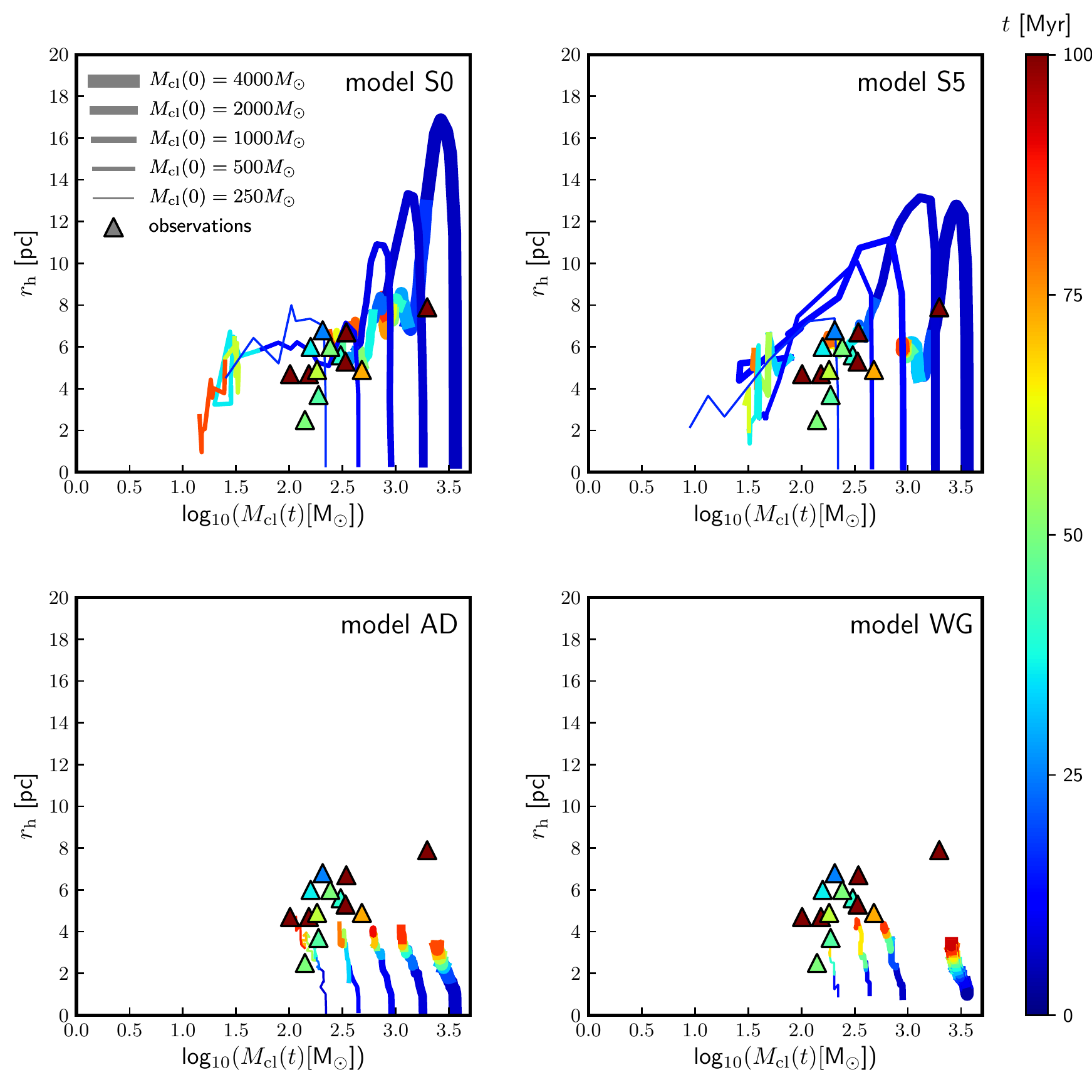}
\caption{
        Evolution of the cluster half-mass radius $r_{\rm h}$ versus the cluster mass $M_{\rm cl}$ in $N$-body simulations. 
         Each panel features a different set of cluster initial conditions as indicated by the model name at the upper-right corner of each panel. 
        The evolution of clusters is shown with the curves, where the thickness of the curve represents the initial cluster mass, and its colour represents the age (shown in the colourbar). The observed 13 target clusters are represented with triangles, again with colors indicating their ages. 
        Note that models~S0 and~S5 are consistent with most of the observational data (particularly with clusters of $M_{\rm cl} \gtrsim 250\,{\rm M}_\odot$), 
        while models~AD are consistent only with clusters of $M_{\rm cl} \lesssim 250\,{\rm M}_\odot$. Models~WG are largely inconsistent with most observed clusters.	     
	    }
\label{fig:rh_mass_nbody}
\end{figure*}

Figure~\ref{fig:rh_mass_nbody} shows the relationship between the cluster's half-mass radius $r_{\rm h}$, its total mass $M_{\rm cl}$ and its age $t$ (color-scheme). Immediately after gas expulsion, models~S0 and~S5 increase their $r_{\rm h}$ substantially. They revirialize, at which stage both $r_{\rm h}$ and $M_{\rm cl}$ decrease. Primordially mass-segregated clusters (model~S5) expand somewhat less. The evolution of cluster mass and half-mass radius for both models~S0 and~S5 is in agreement with the majority of the target clusters (triangles), where only two clusters (IC~2391 and IC~2602) have their radii too compact for their age and mass. In contrast, the properties of models AD and WG are inconsistent with many of the target clusters.

The agreement between the models S0 and S5 with the target clusters more massive than $\log_{10}(M_{cl}(t))\ga2.4$ suggests that these clusters formed with a relatively low SFE (around 1/3), and with a process gas expulsion operating on a time-scale shorter than the stellar crossing time. This finding was also presented in earlier work \citep{Banerjee2017} for more massive star clusters ($\log_{10}(M_{cl}(t)) \gtrsim 4$), extending their result to clusters down to mass $\approx 250\,{\rm M}_\odot$. 
Another piece of evidence for rapid gas expulsion in these clusters is the higher value of dynamical mass as compared to the photometric mass. 

As mentioned above, the state of the two most compact clusters (IC~2391 and IC~2602) appears to be in disagreement with models S0 and S5. However, these two clusters are consistent with models AD. It is impossible to draw a firm conclusion from this based only on two star clusters, but the data may suggest that the gas expulsion time-scale transitions from adiabatic to impulsive at cluster mass of $\approx 250\,{\rm M}_\odot$, while the SFE does not change substantially. A decrease of the gas expulsion time-scale with cluster mass is expected theoretically because the maximum stellar mass in a cluster increases with the mass of the cluster \citep[see][and references therein]{Weidner2013}, so that the total energy of photoionising feedback of the cluster increases with cluster mass. The reduction of the gas expulsion time-scale with increasing cluster mass is also reported in hydrodynamic simulations of \citet{dinnbier2020c} and the observational analysis carried out by \citet{Pfalzner2020}.

%fig4

\subsection{Mass segregation}\label{sec:mass_seg}

Mass segregation is commonly found in embedded clusters and young star clusters, and can be a consequence of internal dynamical relaxation, violent relaxation and/or primordial mass segregation \citep{hillenbrand1998,allison2009,pang2013,pavlik2019}. The youngest cluster in our targets, NGC\,2232 does not manifest any evidence of mass segregation, based on measurements of the mean mass in different annuli \citep{pang2020}. 
Two mediate-age clusters in our sample, however, do show mass segregation. Coma Berenices shows evidence of mass segregation that was quantified by comparing the mass distributions in different annuli \citep{tang2018}, and Blanco\,1 shows evidence of mass segregation obtained using the $\Lambda$-method \citep{zhang2020}.

The $\Lambda$-method, developed by \citet{allison2009}, is a tool to analyse the degree of mass segregation a star cluster without the necessity of determining of cluster center. The $\Lambda$-method compares the minimum path length among the $N_{\rm massive}$ most massive members ($l_{\rm massive}$) of the cluster, to that of the minimum path length of $N_{\rm normal}$ random members ($l_{\rm normal}$). 

This average minimum path length $l$ is calculated from the minimum spanning tree (MST) of the sample of stars, which is obtained using the Python package \texttt{MiSTree} \citep{naidoo2019}.
When the $N_{\rm massive}$ massive stars are segregated, the average path length for this set of stars, $l_{\rm massive}$, is smaller than that for the set of randomly selected stars ($l_{\rm normal}$).  

Previous studies have applied the $\Lambda$-method to star clusters using the observed 2D positions of stars in the clusters. Examples include the studies of NGC\,3603 in \citet{pang2013} and of Blanco\,1 in \citet{zhang2020}. However, the 2D projection can overestimate the degree of segregation by projecting background stars that are located behind the cluster center into the inner region. With the distance-corrected 3D spatial positions of the target cluster members, we are able to improve the determination of the degree of mass segregation, in 3D space.
The significance of the mass segregation is measured using the ``mass segregation ratio'' 
\citep[$\Lambda_{\rm MSR}$, ][]{allison2009}, which defined as
    \begin{equation}\label{eq:MSR}
        \Lambda_{\rm MSR} = \frac{\langle l_{\rm normal} \rangle}{l_{\rm massive}} 
        \pm \frac{\sigma_{\rm normal}}{l_{\rm massive}} \quad , 
    \end{equation}

where $\sigma_{\rm normal}$ is the standard deviation of the 100 different sets of $l_{\rm normal}$, and $\langle l_{\rm normal} \rangle$ is the average length of a hundred random sets.

\begin{figure*}[tb!]
\centering
\includegraphics[angle=0, width=1.\textwidth]{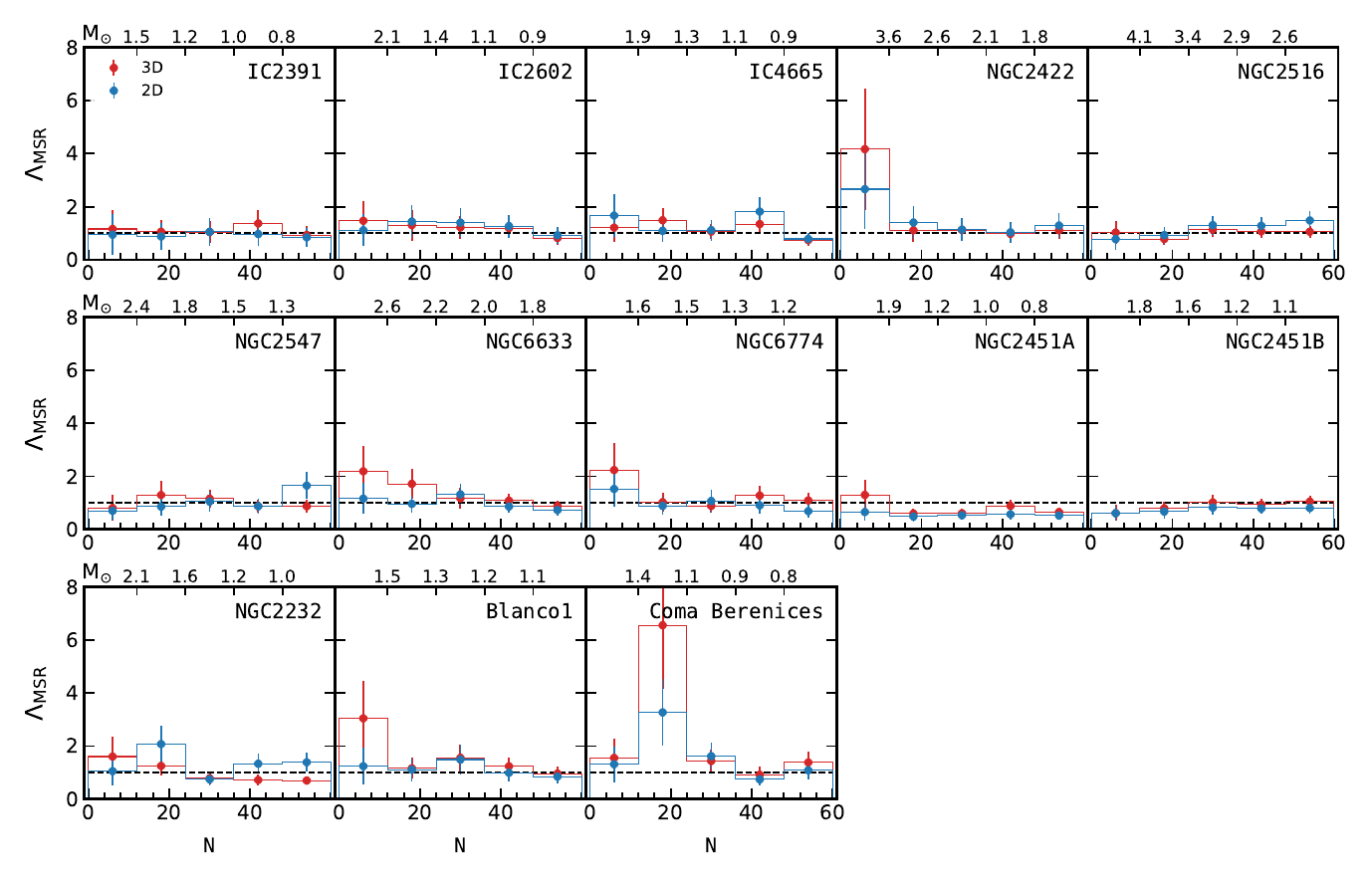}
\caption{
        The ``mass segregation ratio'' ($\Lambda_{\rm MST}$) for the 60 most 
        massive members, with a bin size of 12 stars in each target cluster.
        The dashed line ($\Lambda_{\rm MST} = 1$) indicates an absence of mass segregation. Increasing value of $\Lambda_{\rm MST}$ indicate a more significant degree of mass segregation. The error bars indicate the uncertainties obtained from a hundred realizations of $\emph{l}_{\rm normal}$. The bin size is selected to avoid large stochastic errors in  $\langle l_{\rm normal} \rangle$ for small $N_{\rm MST}$.}
\label{fig:mst}
\end{figure*}

Figure~\ref{fig:mst} presents the $\Lambda_{\rm MSR}$ for the clusters, based on the 3D and 2D positions of members in each cluster. As can be seen from the figure, robust evidence of mass segregation is found in six clusters, NGC\,2422 (segregated down to 3.6 M$_\odot$), NGC\,6633 (2.2 M$_\odot$), NGC\,6774 (1.6 M$_\odot$), NGC\,2232 (2.1 M$_\odot$), Blanco\,1 (1.5 M$_\odot$) and Coma Berenices (1.1~M$_\odot$), consistent with previous works \citep{prinstinzao2003,kraus2007,moraux2007,tang2018,yeh2019}. In Coma Berenices, the most massive stars are not the most concentrated; they have likely been expelled of the cluster center via close encounter with binary stars \citep[e.g.,][]{oh2015,oh2018}. 
No evidence for mass segregation is found for the other seven target clusters. 

The 2D projected distance will reduce the value of $\langle l_{\rm normal} \rangle$ in Equation~\ref{eq:MSR} by projecting stars that are located further away from cluster center onto the inner region. This will result in a decrease in $\Lambda_{\rm MSR}$. Therefore, the 2D MST will most likely under-estimate the degree of mass segregation in a star cluster (Figure~\ref{fig:mst}).

%fig4

\section{Summary}\label{sec:summary}

Utilizing high-precision {\it Gaia} EDR\,3 astrometry and photometry, we apply the cluster finding method
\textsc{StarGO} to identify member stars in 13 target clusters: IC\,2391, IC\,2602, IC\,4665, NGC\,2422, NGC\,2516, NGC\,2547, NGC\,6633, NGC\,6774, NGC\,2451A, and NGC\,2451B, NGC\,2232, Blanco\,1, and Coma Berenices in the 5D phase space of stars ($X, Y, Z$, $\mu_\alpha \cos\delta, \mu_\delta$). The selected members are cross-matched with members in catalogs of \citet{cantat2020} and \citet{liu2019}. The ages obtained from isochrone fitting for each cluster agree with those of previous studies. Altogether we have 13 target clusters with members determined via the same method, covering an age range from 25\,Myr to 2.65\,Gyr, and located in the solar neighbourhood up to a distance of 500\,pc. We analyze the 3D morphology and cluster dynamics of these 13 clusters, and quantify their morphology and dynamical state. $N$-body simulations are carried out to determine which gas expulsion scenario best describes the history of these star clusters. Our findings can be summarized as follows.

%-------------------------------------------------------------------------------------------%
\begin{enumerate}

\item We recovered the individual distance of each candidate member from the parallax by means of a Bayesian method.
The uncertainties in the corrected distances are estimated by simulations of spherical clusters with a uniform spatial distribution of members, and of clusters with elongated shapes. The estimated distance for a uniform-density, spherical model has an uncertainty of 3.0\,pc in the distance when the cluster is located at 500\,pc. Elongated models suffer from larger uncertainty. Notably, when the elongation is along the line-of-sight, uncertainties in the distance reach 6.3\,pc at the distance of 500\,pc.

\item We have determined the 3D morphology of 13 target OCs, with corrected position in the Cartesian heliocentric coordinates ($X$, $Y$, and $Z$). An ellipsoid model is chosen to fit the spatial distribution of the stars within the tidal radius in all clusters. The semi-major axis $a$, semi-intermediate axis $b$, and semi-minor axis $c$ of the ellipsoid are obtained from fitting. We use the axes lengths $a$, $b$, $c$, and axis ratios $b/a$ and $c/a$ as morphological parameters to quantify the 3D distribution of the stellar population within the tidal radii of the OCs. We consider the direction of the major axis as the direction of the morphological elongation of each cluster. Most clusters have semi-major axes $a$ parallel to the Galactic plane or slightly inclined with respect to the Galactic plane. A notable exception is Blanco\,1, for which $a$ is closer to the vertical ($Z$) direction. The shapes of the distribution of the stellar population within the tidal radius for five clusters (NGC\,2547, NGC\,2516, NGC\,2451A, NGC\,2451B, and NGC\,2232) resemble that of an oblate spheroid, while those of other five clusters (IC\,2602, IC\,4665, NGC\,2422, Blanco\,1 and Coma Berenices) resemble prolate spheroids. The shape of the stellar population within the tidal radii of the other three clusters (IC\,2391, NGC\,6633, NGC\,6774) are well-described by triaxial ellipsoids.

\item A significant elongation is observed for the bound regions of NGC\,2422, NGC\,2457, NGC\,6633 and Blanco\,1. Considering that the uncertainty in the corrected distance is much smaller than the size of elongated structures, the elongations measured for these clusters are robust. Among these, Blanco\,1 is notable in the sense that its elongated shape is significantly inclined (by $78^\circ$) with respect to the Galactic plane. The elongation of the bound region might be driven by evaporation of stars via the two Lagrange points. The 3D morphology of Blanco\,1 might be a result of expansion due to fast gas expulsion and virialisation. Elongated filament-like substructures are found in three young clusters, NGC\,2232, NGC\,2547 and NGC\,2451B, while tidal-tail-like substructures are found in older clusters NGC\,2516, NGC\,6633, NGC\,6774. Giant tidal tails are again confimed in Blanco\,1 and Coma Berenices with {\it Gaia} EDR\,3. 

\item We combine {\it Gaia} EDR\,3 PMs and RVs, together with RVs from \citet{jackson2020} and  \citet{bailey2018} to measure the 3D velocity of stellar members in the 13 target clusters. All clusters show evidence of expansion in their 3D velocity distributions. There is an anisotropy in the velocity dispersion for stars inside the tidal radius, which may be driven by gas expulsion. 

\item Four models of $N$-body simulations are carried out to determine the properties of the the gas expulsion process that have occurred in the target clusters: (i) a non mass-segregated model with impulsive gas expulsion; (ii) a mass-segregated model with impulsive gas expulsion; (iii) a model with adiabatic gas expulsion; and (iv) a model without gas. All the target clusters with a mass larger than 250\,M$_\odot$ are consistent with models of rapid (impulsive) gas expulsion with a rather low SFE of $\approx 1/3$, for models both with and without primordial mass segregation. The target clusters with mass smaller than 250\,M$_\odot$ are consistent with models of slow (adiabatic) gas expulsion with an SFE of $\approx 1/3$. Although the results for clusters with masses above 250\,M$_\odot$ appear to be robust, the results for lower mass clusters are only tentative as they are based only on a sample of two clusters. If the decrease of gas expulsion time-scale with increasing cluster mass is confirmed for more clusters in the future, this may point towards a prominent role of feedback from massive stars on the early evolution of star clusters. Models without gas expulsion, i.e., models assuming a star formation efficiency of 100~percent, are not compatible with the data.

\item In order to quantify the degree of mass segregation in each cluster, we apply both 3D and 2D MST methods to the OCs in the sample. Six of the OCs in our sample are found to have mass segregation: NGC\,2422, NGC\,6633, and NGC\,6774, NGC\,2232, Blanco\,1, and Coma Berenices. 
\end{enumerate}
Our study of these 13 open clusters is a pioneering attempt in quantitative study of cluster morphology and its relation to the formation and early evolution of star clusters. The methods developed in this work can be applied to study a much larger sample of OCs covering different locations with data from the {\it Gaia} EDR\,3 and DR\,3, with the aim of achieving a better explanation of the dependence of 3D morphology of open clusters on the location of the star clusters in the Galaxy.

%-------------------------------------------------------------------------------------------%
%-------------------------------------------------------------------------------------------%
\acknowledgments
We wish to express our gratitude to the anonymous referee for providing comments and suggestions that helped to improve the quality of this paper. 
X.Y.P. is grateful to the financial support of the research development fund of Xi'an 
Jiaotong-Liverpool University (RDF-18--02--32). This study is supported by XJTLU Undergraduate Summer Internship in Physics (X-SIP). X.Y.P. gave thanks two grants of National Natural Science Foundation of China, No: 11503015 and 11673032. M.B.N.K. expresses gratitude to the National Natural Science Foundation of China (grant No. 11573004) and the Research Development Fund (grant RDF-16--01--16) of Xi'an Jiaotong-Liverpool University (XJTLU). Franti\v{s}ek Dinnbier and Pavel Kroupa acknowledge support from the Grant Agency of the Czech Republic under grant number 20-21855S as well as support through the DAAD East-European partnership exchange programme.
M.P.'s contribution to this material is based upon work supported by Tamkeen under the NYU Abu Dhabi Research Institute grant CAP$^3$.

This work made use of data from the European Space Agency (ESA) mission {\it Gaia} 
(\url{https://www.cosmos.esa.int/gaia}), processed by the {\it Gaia} Data Processing 
and Analysis Consortium 
(DPAC, \url{https://www.cosmos.esa.int/web/gaia/dpac/consortium}). This study also made use of 
the SIMBAD database and the VizieR catalogue access tool, both operated at CDS, Strasbourg, France.

%------------------

\software{  \texttt{Astropy} \citep{astropy2013,astropy2018}, 
            \texttt{SciPy} \citep{millman2011},
            \texttt{TOPCAT} \citep{taylor2005}, and 
            \textsc{StarGO} \citep{yuan2018}
}
%-------------------
%%%%%%%%%%%%%%%%%%%%%%%%%%%%%%%%%%%%%%%%%%%%%%%%%%%%%%%
\clearpage
%-------------------
%%%%%%%%%%%%%%%%%%%%%%%%%%%%%%%%%%%%%%%%%%%%%%%%%%%%%%%
{}

%-------------------------------------------------------------------------------------------%
\appendix{}

\section{Cartesian Galactocentric/Heliocentric coordinates used in this study}\label{sec:apx_coordinate}
The Galactic center ($l=0\degr$ and $b=0\degr$) is located at the origin of the Cartesian Galactocentric coordinate system. 
The Sun is located 27\,pc above the Galactic midplane, and 8.3\,kpc from the Galactic center \citep{chen2001, gillessen2009}.
The positive $X$-axis points from the projection of the Sun's position onto the Galactic mid-plane towards the Galactic center.
The positive $Y$-axis points towards $l=90\degr$, and the positive $Z$--axis points towards $b=90\degr$. The origin of the Cartesian heliocentric coordinate system is the solar system barycenter, while the direction of axes remains unchanged.

\newpage
\section{Figures for Section~2}\label{sec:apx_som}

 %figa.1
\begin{figure*}[hb!]
\centering
\includegraphics[angle=0, width=0.9\textwidth]{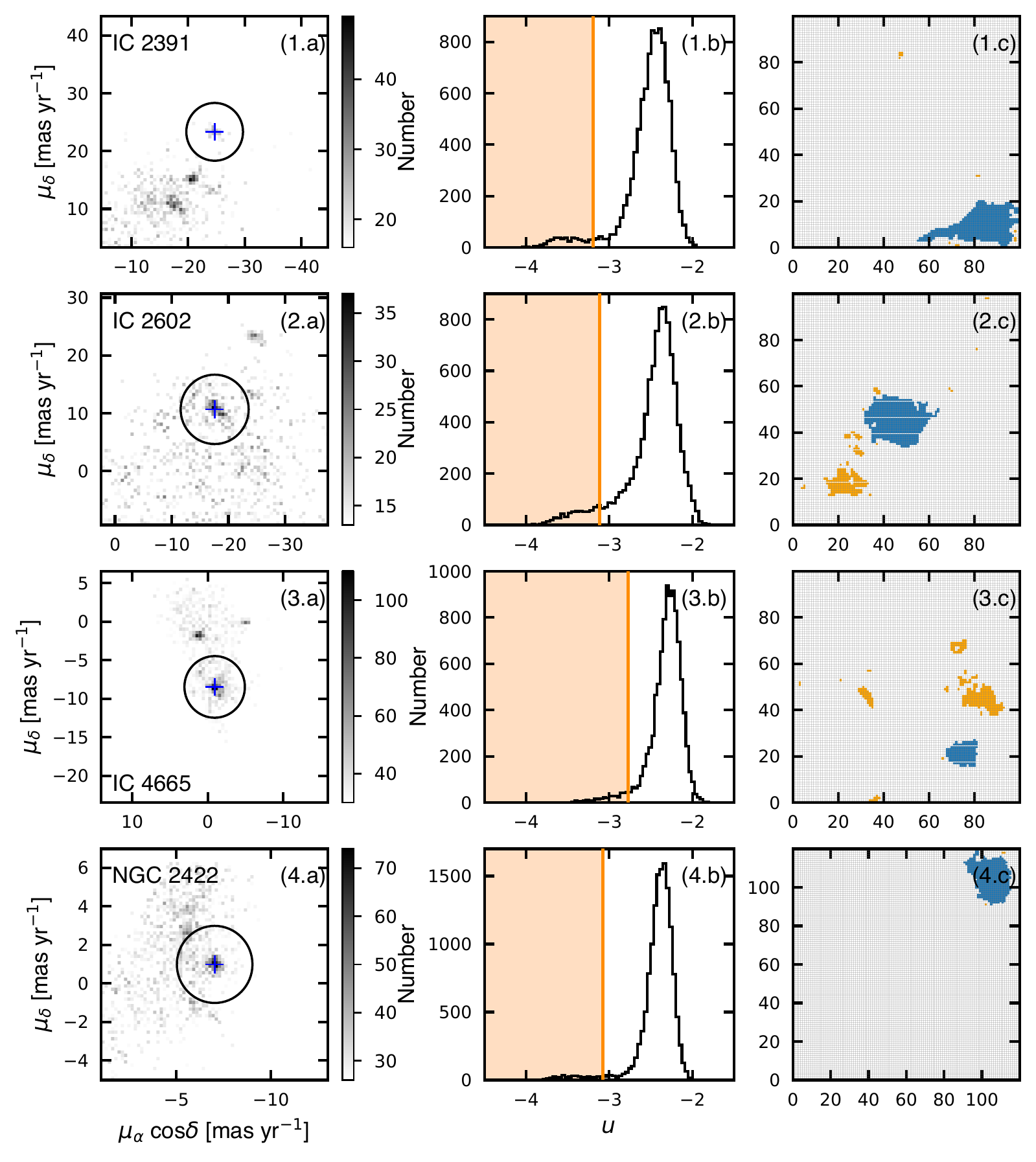}
\caption{(a)~ 2D density map of the proper motion vectors for the regions around four target clusters in sample~I. The blue crosses indicate the mean over-densities generated by the target clusters taken from \citet{liu2019}. Each bin is smoothed by neighboring 8 bins, and here only bins with a number count $>3\sigma$ are shown, where $\sigma$ is the standard deviation of all bins. The greyscale indicates the number count in each bin.
	(b)~ Histogram of the distribution of $u$. 
	    The orange line denotes the selections of $u$ that produces a 5\%
	    contamination rate among the identified candidates, for the orange patch
	    in the 2D neural network (panel (c)).
	(c)~ 2D neural network resulting from SOM, the neurons with a $u$-selection of 5\% contamination rate (orange line in panel (b)) are shaded in orange. Among these, the neurons corresponding to the member candidates of the target cluster are highlighted in blue.  
	    }
\label{fig:som_a1}
\end{figure*}

% fig a.2
\begin{figure*}[tb!]
\centering
\includegraphics[angle=0, width=0.9\textwidth]{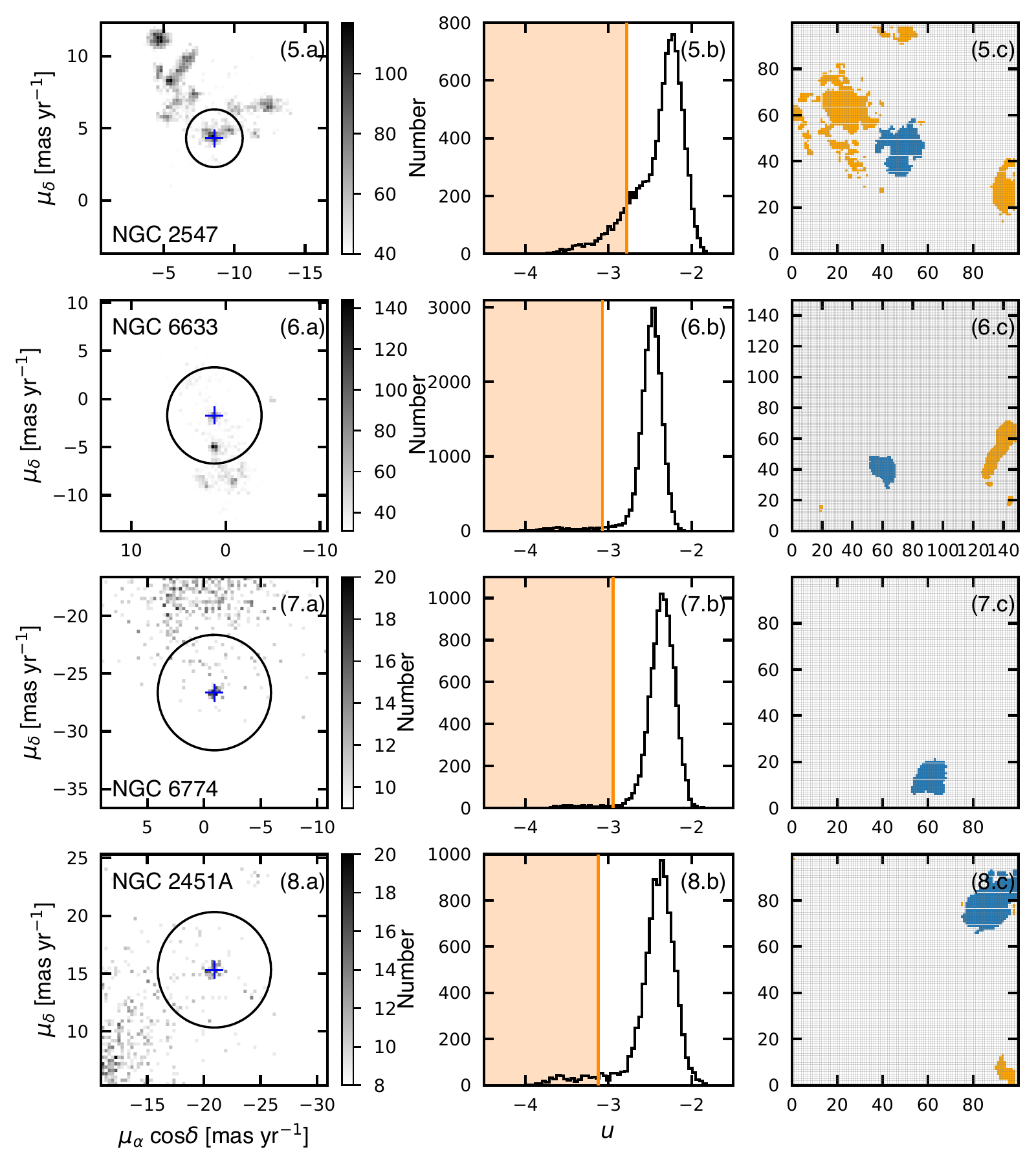}
\caption{ 	 
    (a)~ the 2D density map of the proper motion vectors for the regions around five target clusters in sample~I. 
	(b)~ Histogram of the distribution of $u$. 
	(c)~ 2D neural network resulting from SOM. The symbols and color coding are identical to those in Figure~\ref{fig:som_a1}.
	    }
\label{fig:som_a2}
\end{figure*}

% fig a.2
\begin{figure*}[tb!]
\centering
\includegraphics[angle=0, width=0.9\textwidth]{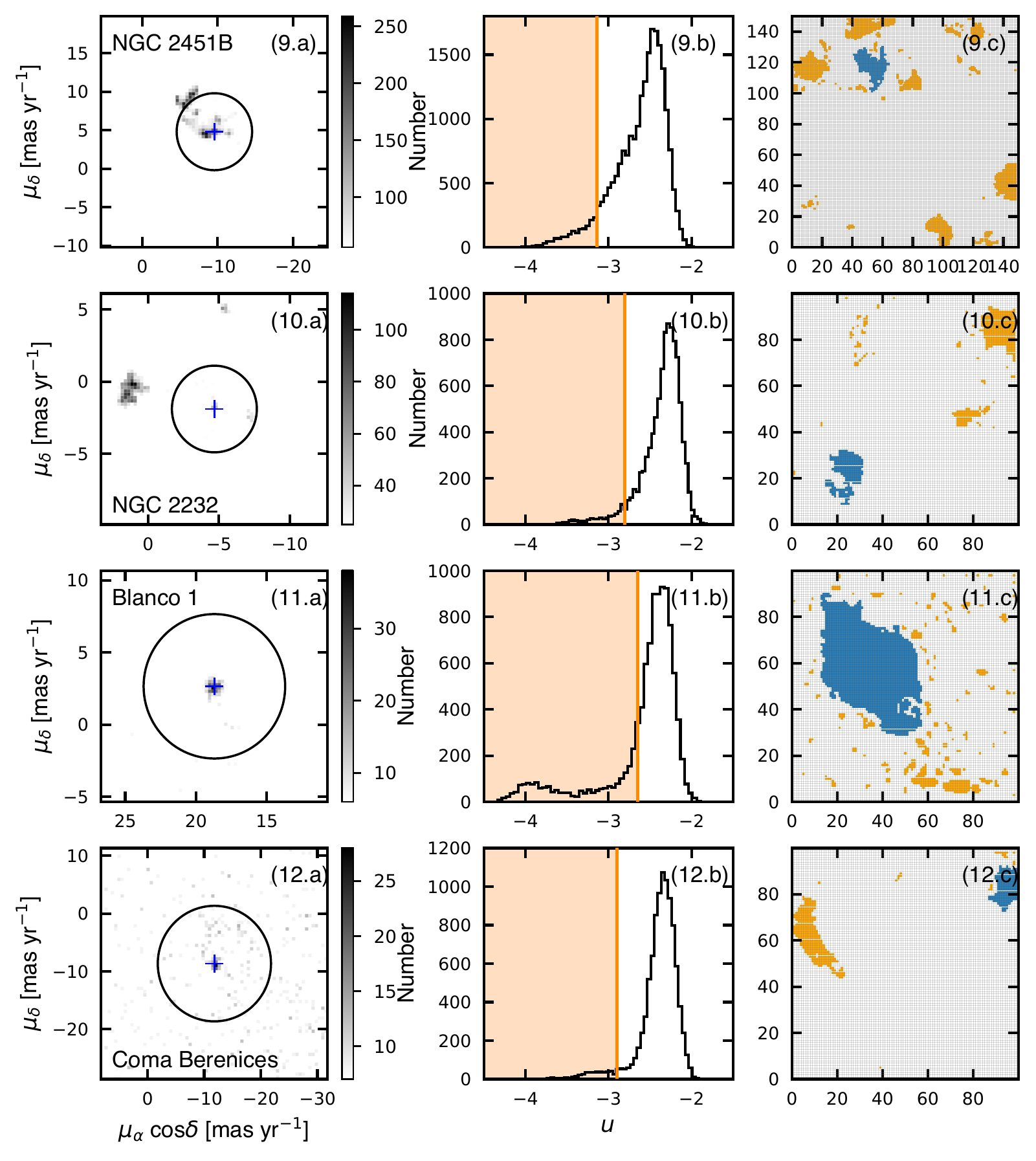}
\caption{ 	 
    (a)~ the 2D density map of the proper motion vectors for the regions around five target clusters in sample~I. 
	(b)~ Histogram of the distribution of $u$. 
	(c)~ 2D neural network resulting from SOM. The symbols and color coding are identical to those in Figure~\ref{fig:som_a1}.}
\label{fig:som_a3}
\end{figure*}

\clearpage

\section{Estimation of uncertainty in the corrected distance}\label{sec:dis_correct_er}

In this section we describe Monte-Carlo simulations that were carried out to quantify the uncertainty in the corrected distances to the individual stars using Bayesian method. In step~I, we generate simulated clusters to model observations. A thousand stars are uniformly distributed within a radius of 10\,pc. The cluster is first placed at a distance of 50\,pc. An initial parallax (hereafter parallax~(I), in step~I) is assigned to each star through inverting its original distance. 
To simulate the observed parallax errors, we resample the observed parallax (I) from a Gaussian distribution, with the initial parallax as the mean and the mean parallax error among members of all clusters  (0.046\,mas~yr$^{-1}$) as the standard deviation. The observed parallax~(I) is converted into observed distance~(I) by reciprocation. An artificial elongated cluster is generated by stretching the stellar population along the line-of-sight, similar to observations. We apply the Bayesian method to correct the observed distances of the individual stars. To maintain consistency with the membership determination applied in Section~\ref{sec:stargo}, we adopt a membership probability of 95\% for each star. The difference between the corrected distance~(I) and the original distance is adopted as the uncertainty of the Bayesian method. We increase the distance of the simulated cluster with a steps of 50\,pc until a distance of 500\,pc. We repeat this procedure for an ensemble of 100 simulations in order to obtain a statistically reliable result. 
%DR2 mean parallax error: 0.078\,mas~yr$^{-1}$)

OCs can be intrinsically elongated. To further investigate the dependence of the distance correction on the intrinsic morphology of star clusters, we also
generate OCs that are intrinsically elongated. To simplify the procedure, we consider the artificially elongated cluster resulting from step~I as a starting point in step~II. In this case, the elongated shape is considered as the original morphology of cluster. Two types of intrinsically elongated clusters are simulated: (i) clusters with an elongation along the line-of-sight and (ii) clusters with an elongation perpendicular to the line-of-sight. We obtain the initial parallax~(II), observed parallax~(II), observed distance~(II), and corrected distance~(II) following the same procedure as in step~I. The elongated simulated clusters are located at distances ranging from 50\,pc to 500\,pc from the Sun. The uncertainty in the corrected distances for the elongated clusters (dotted and dashed curves in Figure~\ref{fig:simulation_dis_er}) follow similar trend as the those of the uniform cluster. 

In general, intrinsically elongated clusters have larger uncertainties in their corrected distances than clusters with a spherical  uniform   stellar distribution. When the cluster is elongated perpendicular to the line-of-sight, the uncertainty in the corrected distance is close to that of a uniform model for distances smaller than approximately 300\,pc. At distances larger than 300\,pc, errors in elongated models are larger than those of uniform cluster 3.0\,pc, and reach 3.4\,pc at 500\,pc. The situation is different in the model with elongation along the line-of-sight. For distances larger than 200\,pc, such clusters show significant deviations in the uncertainty of the corrected distance when compared to the uniform model, and reach an uncertainty of 6.3\,pc at a distance of 500\,pc.

\newpage
\section{Ellipsoid fitting}\label{sec:apx_ellipsoid}

% fig a.2
\begin{figure*}[hb!]
\centering
\includegraphics[angle=0, width=0.9\textwidth]{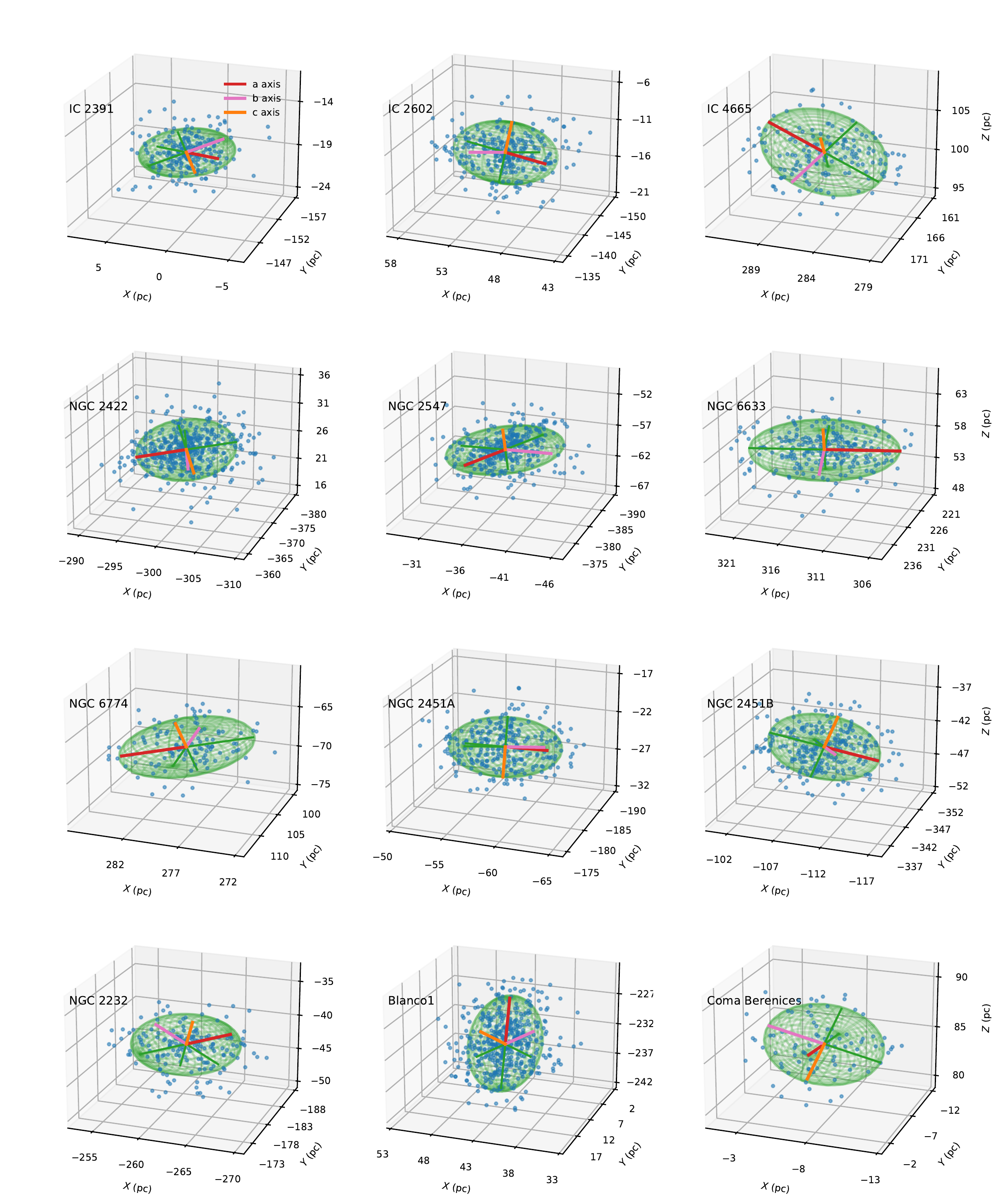}
\caption{ 	 
         Ellipsoid fitting for the 3D spatial positions (in heliocentric Cartesian coordinates $X,Y,Z$) of cluster members within the tidal radii of twelve target clusters after distance correction via a Bayesian approach (see Section~\ref{sec:dis_correct}). The fitted ellipsoid is shown with the green surface. Blue dots are members within tidal radii. The $a$, $b$, and $c$ axes of the  ellipsoid are indicated in red, pink, and orange, respectively. 
	    }
\label{fig:apx_ellip_all}
\end{figure*}

\section{$N$-body simulation setup}\label{sec:apx_nbody}

\subsubsection{Numerical method}

The $N$-body simulations are carried out by the code NBODY6. 
The code uses state-of-the-art numerical techniques \citep{Kustaanheimo1965,Ahmad1973,Aarseth1974a,Mikkola1990,Makino1991,Makino1992} to deal with the large dynamical range of time-steps of the stars under integration.
Stellar evolution and binary evolution algorithms are adopted from  \citet{Tout1996,Hurley2000,Hurley2002}. 
The modelled clusters are subjected to the external gravitational potential of the Galaxy, which is approximated by the model of \citet{Allen1991}. A detailed description of NBODY6 and further applications can be found in \citet{spurzem1999}, \citet{Aarseth2003} and \citet{wang2015, wang2016}. 

\subsubsection{Initial binary conditions}

Binaries of lower mass primary ($m_{\rm break} < m$, where $m_{\rm break} = 5 {\rm M_\odot}$) have orbital parameters 
(i.e. the semi-major axis and eccentricity) and mass ratios generated from the initial binary distribution of \citet{kroupa1995a}, while binaries of the primary more massive than that are generated according to the distribution 
of \citet{Sana2012} and \citet{Moe2017}. 
The initial conditions for the star clusters are generated using the software package \textsc{mcluster} \citep{Kupper2011}.

\subsubsection{Gas expulsion}

The gaseous component of the cluster is approximated with an analytical gravitational potential, which also follows the Plummer profile with the same
half-mass radius as the stellar component. 
The initial mass $M_{\rm gas}(0)$ of the gaseous component is given by the definition of the SFE, which we adopt for simplicity 
as $\mathrm{SFE} = M_{\rm cl}
(0)/(M_{\rm cl}(0) + M_{\rm gas}(0))$. 
The gaseous potential does not evolve up to $t_{\rm d} = 0.6$\,Myr, whereupon it is reduced as 
\begin{equation}
M_{\rm gas}(t) = M_{\rm gas}(0) \exp{\left\{ (t - t_{\rm d})/\tau_{\rm M} \right\}},
\label{eGasPot}
\end{equation}
where $\tau_{\rm M}$ is the gas expulsion time-scale. This follows the procedure of \citet{kroupa2001} and the conditions for the ultra compact HII region phase. 

For each set of model, we obtain a realization of the most massive cluster twice with a different random number seed, and cluster of initial mass $M_{\rm cl}(0) = 2^{-i} \times 4000 \;{\rm M_\odot}$ are realised $2^{i+1}$ times. All the results are then averaged over the ensemble of models realised with  different random number seeds. 
%*******************************************************************************************%
%-------------------------------------------------------------------------------------------%

\begin{thebibliography}{}

\bibitem[{{Aarseth}(2003)}]{Aarseth2003}
        {Aarseth}, S.~J. 2003, {Gravitational N-Body Simulations} (Cambridge: Cambridge University Press)

\bibitem[{{Aarseth} {et~al.}(1974){Aarseth}, {Henon}, \& {Wielen}}]
        {Aarseth1974a} {Aarseth}, S.~J., {Henon}, M., \& {Wielen}, R. 1974, \aap, 37, 183

\bibitem[{{Ahmad} \& {Cohen}(1973)}]{Ahmad1973}
        {Ahmad}, A. \& {Cohen}, L. 1973, Journal of Computational Physics, 12, 389

\bibitem[{{Allen} \& {Santillan}(1991)}]{Allen1991}
        {Allen}, C. \& {Santillan}, A. 1991, \rmxaa, 22, 255

\bibitem[Allison et al.(2009)]{allison2009} 
        Allison, R.~J., Goodwin, S.~P., Parker, R.~J., et al.\ 2009, \apjl, 700, L99. doi:10.1088/0004-637X/700/2/L99

\bibitem[Astropy Collaboration et al.(2013)]{astropy2013} 
            Astropy Collaboration, Robitaille, T.~P., Tollerud, E.~J., et al.\ 2013, \aap, 558, A33

\bibitem[Astropy Collaboration et al.(2018)]{astropy2018} 
            Astropy Collaboration, Price-Whelan, A.~M., Sip{\H o}cz, B.~M., et al.\ 2018, \aj, 156, 123

\bibitem[Bailer-Jones(2015)]{bailer2015} 
            Bailer-Jones, C.~A.~L.\ 2015, \pasp, 127, 994

\bibitem[Bailey et al.(2018)]{bailey2018} 
        Bailey, J.~I., Mateo, M., White, R.~J., et al.\ 2018, \mnras, 475, 1609. doi:10.1093/mnras/stx3266

\bibitem[Ballone et al.(2020)]{ballone2020} 
        Ballone, A., Mapelli, M., Di Carlo, U.~N., et al.\ 2020, \mnras, 496, 49. doi:10.1093/mnras/staa1383

\bibitem[Balog et al.(2009)]{balog2009} 
        Balog, Z., Kiss, L.~L., Vink{\'o}, J., et al.\ 2009, \apj, 698, 1989. doi:10.1088/0004-637X/698/2/1989

\bibitem[Bate(2012)]{bate2012} 
        Bate, M.~R.\ 2012, \mnras, 419, 3115. doi:10.1111/j.1365-2966.2011.19955.x

\bibitem[\protect\citeauthoryear{{Banerjee} \& {Kroupa}}{{Banerjee} \& {Kroupa}}{2017}]{Banerjee2017}
        {Banerjee} S., {Kroupa} P., 2017, \aap, 597, A28

\bibitem[{{Banerjee} \& {Kroupa}(2018)}]{Banerjee2018} 
        {Banerjee}, S. \& {Kroupa}, P. 2018, {Formation of Very Young Massive Clusters and Implications for Globular Clusters}, ed. S.~{Stahler}, Vol. 424, 143

\bibitem[Bastian \& de Mink(2009)]{bastian2009} 
        Bastian, N. \& de Mink, S.~E.\ 2009, \mnras, 398, L11. doi:10.1111/j.1745-3933.2009.00696.x

\bibitem[Baumgardt \& Kroupa(2007)]{baumgardt2007} 
        Baumgardt, H., \& Kroupa, P.\ 2007, \mnras, 380, 1589

\bibitem[Beccari et al.(2020)]{beccari2020} 
        Beccari, G., Boffin, H.~M.~J., \& Jerabkova, T.\ 2020, \mnras, 491, 2205

\bibitem[Belloni et al.(2017)]{belloni2017} 
        Belloni, D., Askar, A., Giersz, M., et al.\ 2017, \mnras, 471, 2812. doi:10.1093/mnras/stx1763

\bibitem[Benacchio \& Galletta(1980)]{benacchio1980} 
        Benacchio, L. \& Galletta, G.\ 1980, \mnras, 193, 885. doi:10.1093/mnras/193.4.885

\bibitem[Bergond et al.(2001)]{bergond2001} 
        Bergond, G., Leon, S., \& Guibert, J.\ 2001, \aap, 377, 462. doi:10.1051/0004-6361:20011043

\bibitem[Bianchini et al.(2018)]{bianchini2018} 
        Bianchini, P., van der Marel, R.~P., del Pino, A., et al.\ 2018, \mnras, 481, 2125. doi:10.1093/mnras/sty2365

% \bibitem[Boily \& Kroupa(2003)]{boily2003} Boily, C.~M., \& Kroupa, P.\ 2003, \mnras, 338, 665

\bibitem[Boubert et al.(2019)]{boubert2019} 
        Boubert, D., Strader, J., Aguado, D., et al.\ 2019, \mnras, 486, 2618. doi:10.1093/mnras/stz253

\bibitem[Bovy(2017)]{bovy2017} Bovy, J.\ 2017, \mnras, 468, L63

\bibitem[Brandner(2008)]{brandner2008} Brandner, W.\ 2008, arXiv:0803.1974

\bibitem[Bravi et al.(2018)]{bravi2018} 
        Bravi, L., Zari, E., Sacco, G.~G., et al.\ 2018, \aap, 615, A37

\bibitem[Cantat-Gaudin et al.(2018)]{cantat2018} Cantat-Gaudin, T., Jordi, C., Vallenari, A., et al.\ 2018, \aap, 618, A93.

\bibitem[Cantat-Gaudin et al.(2020)]{cantat2020} 
        Cantat-Gaudin, T., Anders, F., Castro-Ginard, A., et al.\ 2020, \aap, 640, A1. doi:10.1051/0004-6361/202038192

\bibitem[Cantat-Gaudin \& Anders(2020)]{cantat2020a} 
        Cantat-Gaudin, T. \& Anders, F.\ 2020, \aap, 633, A99. doi:10.1051/0004-6361/201936691

\bibitem[Castro-Ginard et al.(2018)]{castro2018} 
        Castro-Ginard, A., Jordi, C., Luri, X., et al.\ 2018, \aap, 618, A59. doi:10.1051/0004-6361/201833390

\bibitem[Castro-Ginard et al.(2019)]{castro2019} 
        Castro-Ginard, A., Jordi, C., Luri, X., et al.\ 2019, \aap, 627, A35. doi:10.1051/0004-6361/201935531

\bibitem[Castro-Ginard et al.(2020)]{castro2020} 
        Castro-Ginard, A., Jordi, C., Luri, X., et al.\ 2020, \aap, 635, A45. doi:10.1051/0004-6361/201937386

\bibitem[Carrera et al.(2019)]{carrera2019} 
        Carrera, R., Pasquato, M., Vallenari, A., et al.\ 2019, \aap, 627, A119

\bibitem[Chen et al.(2004)]{chen2004} 
        Chen, W.~P., Chen, C.~W., \& Shu, C.~G.\ 2004, \aj, 128, 2306. doi:10.1086/424855

\bibitem[Chen et al.(2001)]{chen2001} 
        Chen, B., Stoughton, C., Smith, J.~A., et al.\ 2001, \apj, 553, 184.
        
\bibitem[Cottaar et al.(2012)]{cottaar2012} 
        Cottaar, M., Meyer, M.~R., \& Parker, R.~J.\ 2012, \aap, 547, A35. doi:10.1051/0004-6361/201219673

% \bibitem[Cropper et al.(2018)]{cropper2018} Cropper, M., Katz, D., Sartoretti, P., et al.\ 2018, \aap, 616, A5

\bibitem[Curry(2002)]{curry2002} 
        Curry, C.~L.\ 2002, \apj, 576, 849. doi:10.1086/341811

\bibitem[D'Antona et al.(2017)]{dantona2017}
        D'Antona, F., Milone, A.~P., Tailo, M., et al.\ 2017, Nature Astronomy, 1, 0186. doi:10.1038/s41550-017-0186

\bibitem[Darma et al.(2019)]{darma2019} 
        Darma, R., Arifyanto, M.~I., \& Kouwenhoven, M.~B.~N.\ 2019, Journal of Physics Conference Series, 1231, 012028. doi:10.1088/1742-6596/1231/1/012028

\bibitem[Dinnbier \& Kroupa(2020a)]{dinnbier2020a} Dinnbier, F. \& Kroupa, P.\ 2020, \aap, 640, A85. doi:10.1051/0004-6361/201936572

\bibitem[Dinnbier \& Kroupa(2020b)]{dinnbier2020b} Dinnbier, F. \& Kroupa, P.\ 2020, \aap, 640, A84. doi:10.1051/0004-6361/201936570

\bibitem[Dinnbier \& Walch(2020)]{dinnbier2020c} Dinnbier, F. \& Walch, S.\ 2020, \mnras, 499, 748. doi:10.1093/mnras/staa2560

\bibitem[Duch{\^e}ne \& Kraus(2013)]{duchene2013} Duch{\^e}ne, G. \& Kraus, A.\ 2013, \araa, 51, 269. doi:10.1146/annurev-astro-081710-102602

\bibitem[Fleck et al.(2006)]{fleck2006} 
        leck, J.-J., Boily, C.~M., Lan{\c{c}}on, A., et al.\ 2006, \mnras, 369, 1392. doi:10.1111/j.1365-2966.2006.10390.x

\bibitem[{{Fujii} \& {Portegies Zwart}(2015)}]{Fujii2015a}
        {Fujii}, M.~S. \& {Portegies Zwart}, S. 2015, \mnras, 449, 726

\bibitem[{{Goodwin} \& {Whitworth}(2004)}]{Goodwin2004}
        {Goodwin}, S.~P. \& {Whitworth}, A.~P. 2004, \aap, 413, 929

\bibitem[Einsel \& Spurzem(1999)]{einsel1999} 
        Einsel, C. \& Spurzem, R.\ 1999, \mnras, 302, 81. doi:10.1046/j.1365-8711.1999.02083.x

\bibitem[F{\"u}rnkranz et al.(2019)]{furnkranz2019} 
        F{\"u}rnkranz, V., Meingast, S., \& Alves, J.\ 2019, \aap, 624, L11. doi:10.1051/0004-6361/201935293

\bibitem[Gaia Collaboration et al.(2020)]{gaia2020} 
        Gaia Collaboration, Brown, A.~G.~A., Vallenari, A., et al.\ 2020, arXiv:2012.01533

\bibitem[Gaia Collaboration et al.(2018b)]{gaia2018b} 
    Gaia Collaboration, Babusiaux, C., van Leeuwen, F., et al.\ 2018, \aap, 616, A10 

\bibitem[Gaia Collaboration et al.(2018a)]{gaia2018a} 
    Gaia Collaboration, Brown, A.~G.~A., Vallenari, A., et al.\ 2018, \aap, 616, A1 
    
% \bibitem[\protect\citeauthoryear{Gelman et al.}{2020}]{2020arXiv201101808G} Gelman A., Vehtari A., Simpson D., Margossian C.~C., Carpenter B., Yao Y., Kennedy L., et al., 2020, arXiv, arXiv:2011.01808

\bibitem[Gentile Fusillo et al.(2019)]{fusillo2019} 
        Gentile Fusillo, N.~P., Tremblay, P.-E., G{\"a}nsicke, B.~T., et al.\ 2019, \mnras, 482, 4570. doi:10.1093/mnras/sty3016

\bibitem[Getman et al.(2018)]{getman2018} 
        Getman, K.~V., Kuhn, M.~A., Feigelson, E.~D., et al.\ 2018, \mnras, 477, 298. doi:10.1093/mnras/sty473
   
\bibitem[Gillessen et al.(2009)]{gillessen2009} 
    Gillessen, S., Eisenhauer, F., Trippe, S., et al.\ 2009, \apj, 692, 1075.
    
\bibitem[Gilmore et al.(2012)]{gilmore2012} 
        Gilmore, G., Randich, S., Asplund, M., et al.\ 2012, The Messenger, 147, 25

\bibitem[Goodwin \& Kroupa(2005)]{goodwin2005} 
        Goodwin, S.~P. \& Kroupa, P.\ 2005, \aap, 439, 565. doi:10.1051/0004-6361:20052654
    
\bibitem[Gonz{\'a}lez-Samaniego \& Vazquez-Semadeni(2020)]{gonzalez2020} 
        Gonz{\'a}lez-Samaniego, A. \& Vazquez-Semadeni, E.\ 2020, \mnras, 499, 668. doi:10.1093/mnras/staa2921
    
\bibitem[Hillenbrand \& Hartmann(1998)]{hillenbrand1998} 
        Hillenbrand, L.~A. \& Hartmann, L.~W.\ 1998, \apj, 492, 540. doi:10.1086/305076

\bibitem[Hong et al.(2013)]{hong2013} 
        Hong, J., Kim, E., Lee, H.~M., et al.\ 2013, \mnras, 430, 2960. doi:10.1093/mnras/stt099

\bibitem[{{Hurley} {et~al.}(2000){Hurley}, {Pols}, \& {Tout}}]{Hurley2000}
        {Hurley}, J.~R., {Pols}, O.~R., \& {Tout}, C.~A. 2000, \mnras, 315, 543

\bibitem[{{Hurley} {et~al.}(2002){Hurley}, {Tout}, \& {Pols}}]{Hurley2002}
        {Hurley}, J.~R., {Tout}, C.~A., \& {Pols}, O.~R. 2002, \mnras, 329, 897

\bibitem[Jackson et al.(2020)]{jackson2020} Jackson, R.~J., Jeffries, R.~D., Wright, N.~J., et al.\ 2020, \mnras, doi:10.1093/mnras/staa1749

\bibitem[Jeans(1916)]{jeans1916} Jeans, J.~H.\ 1916, \mnras, 76, 567. doi:10.1093/mnras/76.7.567

\bibitem[Jerabkova et al.(2019)]{jerabkova2019} 
            Jerabkova, T., Boffin, H.~M.~J., Beccari, G., et al.\ 2019, \mnras, 489, 4418

\bibitem[Jones \& Basu(2002)]{jones2002} 
        Jones, C.~E. \& Basu, S.\ 2002, \apj, 569, 280. doi:10.1086/339230

\bibitem[Kamann et al.(2018)]{kamann2018} 
        Kamann, S., Husser, T.-O., Dreizler, S., et al.\ 2018, \mnras, 473, 5591. doi:10.1093/mnras/stx2719

\bibitem[Karnath et al.(2019)]{karnath2019} 
        Karnath, N., Prchlik, J.~J., Gutermuth, R.~A., et al.\ 2019, \apj, 871, 46. doi:10.3847/1538-4357/aaf4c1

\bibitem[Koester \& Reimers(1996)]{koester1996} 
        Koester, D. \& Reimers, D.\ 1996, \aap, 313, 810

\bibitem[Kounkel \& Covey(2019)]{kounkel2019} Kounkel, M., \& Covey, K.\ 2019, \aj, 158, 122

\bibitem[Kouwenhoven \& de Grijs(2008)]{kouwenhoven2008} 
        Kouwenhoven, M.~B.~N. \& de Grijs, R.\ 2008, \aap, 480, 103. doi:10.1051/0004-6361:20078897

\bibitem[Kraus \& Hillenbrand(2007)]{kraus2007} 
        Kraus, A.~L. \& Hillenbrand, L.~A.\ 2007, \aj, 134, 2340. doi:10.1086/522831

\bibitem[Kroupa(1995a)]{kroupa1995a} Kroupa, P.\ 1995, \mnras, 277, 1491. doi:10.1093/mnras/277.4.1491

\bibitem[Kroupa(1995b)]{kroupa1995b} Kroupa, P.\ 1995, \mnras, 277, 1507. doi:10.1093/mnras/277.4.1507

\bibitem[Kroupa et al.(2001)]{kroupa2001} 
        Kroupa, P., Aarseth, S., \& Hurley, J.\ 2001, \mnras, 321, 699. doi:10.1046/j.1365-8711.2001.04050.x

\bibitem[{{Kroupa}(2001)}]{Kroupa2001a}{Kroupa}, P. 2001, \mnras, 322, 231

\bibitem[Kruijssen et al.(2012)]{kruijssen2012} 
        Kruijssen, J.~M.~D., Maschberger, T., Moeckel, N., et al.\ 2012, \mnras, 419, 841. doi:10.1111/j.1365-2966.2011.19748.x

\bibitem[Krumholz \& Matzner(2009)]{krumkolz2009} 
        Krumholz, M.~R. \& Matzner, C.~D.\ 2009, \apj, 703, 1352. doi:10.1088/0004-637X/703/2/1352

\bibitem[Kuhn et al.(2019)]{kuhn2019} 
        Kuhn, M.~A., Hillenbrand, L.~A., Sills, A., et al.\ 2019, \apj, 870, 32. doi:10.3847/1538-4357/aaef8c

\bibitem[{{Kustaanheimo} \& {Stiefel}(1965)}]{Kustaanheimo1965}
        {Kustaanheimo}, P. \& {Stiefel}, E. 1965, Reine Angew. Math., 218, 204

\bibitem[{{K{\"u}pper} {et~al.}(2011){K{\"u}pper}, {Maschberger}, {Kroupa}, \& {Baumgardt}}]{Kupper2011}
        {K{\"u}pper}, A. H.~W., {Maschberger}, T., {Kroupa}, P., \& {Baumgardt}, H. 2011, \mnras, 417, 2300

\bibitem[K{\"u}pper et al.(2008)]{kupper2008} 
        K{\"u}pper, A.~H.~W., MacLeod, A., \& Heggie, D.~C.\ 2008, \mnras, 387, 1248. doi:10.1111/j.1365-2966.2008.13323.x

\bibitem[Lada \& Lada(2003)]{lada2003} 
        Lada, C.~J., \& Lada, E.~A.\ 2003, \araa, 41, 57 

\bibitem[Lamers et al.(2005)]{lamers2005} 
        Lamers, H.~J.~G.~L.~M., Gieles, M., Bastian, N., et al.\ 2005, \aap, 441, 117. doi:10.1051/0004-6361:20042241

\bibitem[Li et al.(2014)]{li2014} Li, C., de Grijs, R., \& Deng, L.\ 2014, \nat, 516, 367. doi:10.1038/nature13969

\bibitem[Li et al.(2017)]{li2017} Li, C., de Grijs, R., Deng, L., et al.\ 2017, \apj, 844, 119. doi:10.3847/1538-4357/aa7b36

\bibitem[Li et al.(2019)]{li2019} Li, C., Sun, W., de Grijs, R., et al.\ 2019, \apj, 876, 65. doi:10.3847/1538-4357/ab15d2

\bibitem[Lindegren et al.(2018)]{lindegren2018} Lindegren, L., Hern{\'a}ndez, J., Bombrun, A., et al.\ 2018, \aap, 616, A2 
    
\bibitem[Liu \& Pang(2019)]{liu2019} Liu, L., \& Pang, X.\ 2019, \apjs, 245, 32
    
\bibitem[{{Makino}(1991)}]{Makino1991}
        {Makino}, J. 1991, \apj, 369, 200

\bibitem[{{Makino} \& {Aarseth}(1992)}]{Makino1992}
        {Makino}, J. \& {Aarseth}, S.~J. 1992, \pasj, 44, 141

\bibitem[Marsden et al.(2009)]{marsden2009} 
        Marsden, S.~C., Carter, B.~D., \& Donati, J.-F.\ 2009, \mnras, 399, 888. doi:10.1111/j.1365-2966.2009.15319.x

\bibitem[Ma{\'\i}z Apell{\'a}niz \& Weiler(2018)]{mazapellaniz2018} 
        Ma{\'\i}z Apell{\'a}niz, J. \& Weiler, M.\ 2018, \aap, 619, A180. doi:10.1051/0004-6361/201834051

\bibitem[Marks \& Kroupa(2011)]{marks2011} 
        Marks, M. \& Kroupa, P.\ 2011, \mnras, 417, 1702. doi:10.1111/j.1365-2966.2011.19519.x

\bibitem[Marks et al.(2011)]{marks2011b} 
        Marks, M., Kroupa, P., \& Oh, S.\ 2011, \mnras, 417, 1684. doi:10.1111/j.1365-2966.2011.19257.x

\bibitem[{{Marks} \& {Kroupa}(2012)}]{Marks2012}
        {Marks}, M. \& {Kroupa}, P. 2012, \aap, 543, A8

\bibitem[Meingast \& Alves(2019)]{meingast2019} 
        Meingast, S. \& Alves, J.\ 2019, \aap, 621, L3. doi:10.1051/0004-6361/201834622

\bibitem[{{Mikkola} \& {Aarseth}(1990)}]{Mikkola1990}
        {Mikkola}, S. \& {Aarseth}, S.~J. 1990, Celestial Mechanics and Dynamical Astronomy, 47, 375
  
\bibitem[Milone et al.(2018)]{milone2018} 
        Milone, A.~P., Marino, A.~F., Di Criscienzo, M., et al.\ 2018, \mnras, 477, 2640. doi:10.1093/mnras/sty661

\bibitem[Miret-Roig et al.(2019)]{miret2019} 
        Miret-Roig, N., Bouy, H., Olivares, J., et al.\ 2019, \aap, 631, A57. doi:10.1051/0004-6361/201935518

\bibitem[Millman et al.(2011)]{millman2011} 
        Millman, K. J., Aivazis, M..\ 2011, Computing in Science \& Engineering, 13, 2, 9

\bibitem[{{Moe} \& {Di Stefano}(2017)}]{Moe2017}
        {Moe}, M. \& {Di Stefano}, R. 2017, \apjs, 230, 15

\bibitem[{{Moeckel} \& {Bate}(2010)}]{Moeckel2010}
        {Moeckel}, N. \& {Bate}, M.~R. 2010, \mnras, 404, 721

\bibitem[Moraux et al.(2007)]{moraux2007} 
        Moraux, E., Bouvier, J., Stauffer, J.~R., et al.\ 2007, \aap, 471, 499. doi:10.1051/0004-6361:20066308

\bibitem[Naidoo(2019)]{naidoo2019} 
        Naidoo, K.\ 2019, The Journal of Open Source Software, 4, 1721. doi:10.21105/joss.01721

\bibitem[Nilakshi et al.(2002)]{nilakshi2002} 
        Nilakshi, Sagar, R., Pandey, A.~K., et al.\ 2002, \aap, 383, 153. doi:10.1051/0004-6361:20011719

\bibitem[Oh et al.(2015)]{oh2015} 
        Oh, S., Kroupa, P., \& Pflamm-Altenburg, J.\ 2015, \apj, 805, 92. doi:10.1088/0004-637X/805/2/92

%\bibitem[Oh \& Kroupa(2016)]{oh2016} Oh, S. \& Kroupa, P.\ 2016, \aap, 590, A107. doi:10.1051/0004-6361/201628233

\bibitem[Oh \& Kroupa(2018)]{oh2018} 
        Oh, S. \& Kroupa, P.\ 2018, \mnras, 481, 153. doi:10.1093/mnras/sty2245

\bibitem[Olivares et al.(2019)]{olivares2019} 
        Olivares, J., Bouy, H., Sarro, L.~M., et al.\ 2019, \aap, 625, A115. doi:10.1051/0004-6361/201834924

\bibitem[Oort(1979)]{oort1979} Oort, J.~H.\ 1979, \aap, 78, 312

\bibitem[McKee \& Ostriker(1977)]{mckee1977} 
        McKee, C.~F. \& Ostriker, J.~P.\ 1977, \apj, 218, 148. doi:10.1086/155667

\bibitem[Meingast et al.(2020)]{meingast2020} 
        Meingast, S., Alves, J., \& Rottensteiner, A.\ 2020, arXiv:2010.06591

\bibitem[Padilla \& Strauss(2008)]{padilla2008} 
        Padilla, N.~D. \& Strauss, M.~A.\ 2008, \mnras, 388, 1321. doi:10.1111/j.1365-2966.2008.13480.x

\bibitem[Pang et al.(2013)]{pang2013} 
    Pang, X., Grebel, E.~K., Allison, R.~J., et al.\ 2013, \apj, 764, 73 
    
\bibitem[Pang et al.(2018)]{pang2018} 
        Pang, X., Shen, S., \& Shao, Z.\ 2018, \apjl, 868, L9. doi:10.3847/2041-8213/aaedaa

\bibitem[Pang et al.(2020)]{pang2020} 
        Pang, X., Li, Y., Tang, S.-Y., et al.\ 2020, \apjl, 900, L4. doi:10.3847/2041-8213/abad28

\bibitem[Parker \& Goodwin(2009)]{parker2009} 
        Parker, R.~J. \& Goodwin, S.~P.\ 2009, \mnras, 397, 1041. doi:10.1111/j.1365-2966.2009.15037.x

\bibitem[Pavl{\'\i}k et al.(2019)]{pavlik2019} 
        Pavl{\'\i}k, V., Kroupa, P., \& {\v{S}}ubr, L.\ 2019, \aap, 626, A79. doi:10.1051/0004-6361/201834265

\bibitem[{{Pfalzner}(2020)}]{Pfalzner2020}
        {Pfalzner}, S. 2020, in Star Clusters: From the Milky Way to the Early Universe, ed. A.~{Bragaglia}, M.~{Davies}, A.~{Sills}, \& E.~{Vesperini}, Vol. 351, 208--211

\bibitem[Pinfield et al.(1998)]{pinfield1998} 
        Pinfield, D.~J., Jameson, R.~F., \& Hodgkin, S.~T.\ 1998, \mnras, 299, 955 

\bibitem[R{\"o}ser et al.(2019)]{roser2019} 
        R{\"o}ser, S., Schilbach, E., \& Goldman, B.\ 2019, \aap, 621, L2

\bibitem[Postnikova et al.(2020)]{postnikova2020} 
        Postnikova, E.~S., Elsanhoury, W.~H., Sariya, D.~P., et al.\ 2020, Research in Astronomy and Astrophysics, 20, 016. doi:10.1088/1674-4527/20/2/16

\bibitem[Prisinzano et al.(2003)]{prinstinzao2003} 
        Prisinzano, L., Micela, G., Sciortino, S., et al.\ 2003, \aap, 404, 927. doi:10.1051/0004-6361:20030524

\bibitem[Priyatikanto et al.(2016)]{priyatikanto2016} 
        Priyatikanto, R., Kouwenhoven, M.~B.~N., Arifyanto, M.~I., et al.\ 2016, \mnras, 457, 1339. doi:10.1093/mnras/stw060

\bibitem[Raghavan et al.(2010)]{raghavan2010} 
        Raghavan, D., McAlister, H.~A., Henry, T.~J., et al.\ 2010, \apjs, 190, 1. doi:10.1088/0067-0049/190/1/1

\bibitem[Rybizki et al.(2018)]{rybizki2018} 
        Rybizki, J., Demleitner, M., Fouesneau, M., et al.\ 2018, \pasp, 130, 74101.

\bibitem[S{\'a}nchez \& Alfaro(2009)]{Sanchez2009} 
        S{\'a}nchez, N. \& Alfaro, E.~J.\ 2009, \apj, 696, 2086. doi:10.1088/0004-637X/696/2/2086

\bibitem[{{Sana} {et~al.}(2012){Sana}, {de Mink}, {de Koter}, {Langer}, {Evans}, {Gieles}, {Gosset}, {Izzard}, {Le Bouquin}, \&
                {Schneider}}]{Sana2012}
        {Sana}, H., {de Mink}, S.~E., {de Koter}, A., {et~al.} 2012, Science, 337, 444 s

\bibitem[Seabroke et al.(2020)]{seabroke2020} 
        Seabroke, G., Cropper, M., Baker, S., et al.\ 2020, arXiv:2010.16337

\bibitem[{{Sills} {et~al.}(2018){Sills}, {Rieder}, {Scora}, {McCloskey}, \& {Jaffa}}]{Sills2018}
        {Sills}, A., {Rieder}, S., {Scora}, J., {McCloskey}, J., \& {Jaffa}, S. 2018, \mnras, 477, 1903

\bibitem[Spitzer(1958)]{spitzer1958} Spitzer, L.\ 1958, \apj, 127, 17. doi:10.1086/146435

\bibitem[Spurzem(1999)]{spurzem1999} Spurzem, R.\ 1999, Journal of Computational and Applied Mathematics, 109, 407

\bibitem[{{{\v{S}}ubr} {et~al.}(2008){{\v{S}}ubr}, {Kroupa}, \& {Baumgardt}}]{Subr2008}
        {{\v{S}}ubr}, L., {Kroupa}, P., \& {Baumgardt}, H. 2008, \mnras, 385, 1673

\bibitem[Tang et al.(2018)]{tang2018} 
        Tang, S.-Y., Chen, W.~P., Chiang, P.~S., et al.\ 2018, \apj, 862, 106. doi:10.3847/1538-4357/aacb7a
        
\bibitem[Tang et al.(2019)]{tang2019} 
        Tang, S.-Y., Pang, X., Yuan, Z., et al.\ 2019, \apj, 877, 12

\bibitem[Taylor(2005)]{taylor2005} 
        Taylor, M.~B.\ 2005, Astronomical Data Analysis Software and Systems XIV, 29

\bibitem[Tian(2020)]{tian2020} Tian, H.-J.\ 2020, \apj, 904, 196. doi:10.3847/1538-4357/abbf4b

\bibitem[Torra et al.(2020)]{torra2020} Torra, F., Casta{\~n}eda, J., Fabricius, C., et al.\ 2020, arXiv:2012.06420

\bibitem[{{Tout} {et~al.}(1996){Tout}, {Pols}, {Eggleton}, \& {Han}}]{Tout1996}
        {Tout}, C.~A., {Pols}, O.~R., {Eggleton}, P.~P., \& {Han}, Z. 1996, \mnras, 281, 257

\bibitem[Wang et al.(2015)]{wang2015} 
        Wang, L., Spurzem, R., Aarseth, S., et al.\ 2015, \mnras, 450, 4070. doi:10.1093/mnras/stv817

\bibitem[Wang et al.(2016)]{wang2016} 
        Wang, L., Spurzem, R., Aarseth, S., et al.\ 2016, \mnras, 458, 1450. doi:10.1093/mnras/stw274

\bibitem[Weaver et al.(1977)]{weaver1977} 
        Weaver, R., McCray, R., Castor, J., et al.\ 1977, \apj, 218, 377. doi:10.1086/155692

\bibitem[{{Weidner} {et~al.}(2013){Weidner}, {Kroupa}, \& {Pflamm-Altenburg}}]{Weidner2013}
        {Weidner}, C., {Kroupa}, P., \& {Pflamm-Altenburg}, J. 2013, \mnras, 434, 84

\bibitem[Williams \& Bolte(2007)]{williams2007} 
        Williams, K.~A. \& Bolte, M.\ 2007, \aj, 133, 1490. doi:10.1086/511675

\bibitem[Yeh et al.(2019)]{yeh2019} 
        Yeh, F.~C., Carraro, G., Montalto, M., et al.\ 2019, \aj, 157, 115. doi:10.3847/1538-3881/aaff6c

\bibitem[Yuan et al.(2018)]{yuan2018} 
        Yuan, Z., Chang, J., Banerjee, P., et al.\ 2018, \apj, 863, 26

\bibitem[Zhang et al.(2020)]{zhang2020} 
        Zhang, Y., Tang, S.-Y., Chen, W.~P., et al.\ 2020, \apj, 889, 99

\bibitem[Zhong et al.(2019)]{zhong2019} 
        Zhong, J., Chen, L., Kouwenhoven, M.~B.~N., et al.\ 2019, \aap, 624, A34. doi:10.1051/0004-6361/201834334

\end{thebibliography}
\end{document}